\documentclass[11pt,a4paper]{article}
\usepackage{jheppub}
\usepackage{amsmath}
\usepackage{amsfonts}
\usepackage{graphicx}
\usepackage{hyperref}
\usepackage{subcaption}
\usepackage{xcolor}
\usepackage{bbold}
\usepackage{amsmath,amssymb}
\usepackage{float}

\title{Krylov complexity from integrability to chaos}

\author[a,b]{E. Rabinovici,} 

\author[c]{A. S\'{a}nchez-Garrido,}

\author[a]{R. Shir}

\author[c]{and J. Sonner}

\affiliation[a]{Racah Institute of Physics, The Hebrew University, Jerusalem 9190401, Israel}
\affiliation[b]{Institut des Hautes Etudes Scientifiques (IHES),
Le Bois-Marie, 35,route de Chartres, 91440 Bures-sur-Yvette, France}

\affiliation[c]{Department of Theoretical Physics, University of Geneva, 24 quai Ernest-Ansermet, 1214 Gen\`eve 4, Switzerland}  

\emailAdd{eliezer@mail.huji.ac.il}
\emailAdd{Adrian.SanchezGarrido@unige.ch}
\emailAdd{ruth.shir@mail.huji.ac.il}
\emailAdd{Julian.Sonner@unige.ch}

\abstract{We apply a notion of quantum complexity, called ``Krylov complexity", to study the evolution of systems from integrability to chaos. For this  purpose we investigate the integrable XXZ spin chain, enriched with an integrability breaking deformation that allows one to interpolate between integrable and chaotic behavior.
 K-complexity can act as a probe of the integrable or chaotic nature of the underlying system via its late-time saturation value that is suppressed in the integrable phase and increases as the system is driven to the chaotic phase. We furthermore ascribe the (under-)saturation of the late-time bound to the amount of disorder present in the Lanczos sequence, by mapping the complexity evolution to an auxiliary off-diagonal Anderson hopping model. We compare the late-time saturation of K-complexity  in the chaotic phase with that of random matrix ensembles and find that the chaotic system indeed approaches the RMT behavior in the appropriate symmetry class. We investigate the dependence of the results on the two key ingredients of  K-complexity:  the dynamics of the Hamiltonian and the character of the operator whose time dependence is followed.}
\date{July 2022}

\begin{document}

\maketitle

\section{Introduction}
The notion of ``complexity'' is playing an increasingly important role in a number of physical contexts \cite{nielsen2002quantum}, from computational condensed matter all the way to holographic spacetime \cite{Susskind:2018pmk}. As suggested by the colloquial meaning of `complexity', such a quantity should capture the notion of how `complicated' a physical system is. Quantum mechanically such a notion could refer to states, say with respect to some chosen `simple' reference state, or operators, or perhaps some combination thereof. In fact, a particularly natural notion of complexity is associated with the time evolution generated by the Hamiltonian itself. A mathematically precise definition of the complexity of time evolution under a given Hamiltonian is given by Krylov complexity \cite{Parker:2018yvk,Barbon:2019wsy,Rabinovici:2020ryf} or `K-complexity' for short. To date, several aspects of Krylov complexity have been studied in various setups and systems, for example \cite{Jian:2020qpp, Rabinovici:2021qqt, Bhattacharjee:2022vlt, Hornedal:2022pkc, Trigueros:2021rwj, Noh_2021, PhysRevA.105.L010201, Caputa:2022eye,Kar:2021nbm,  PhysRevA.105.062210, PhysRevResearch.4.013041, Bhattacharya:2022gbz, PhysRevD.104.L081702, Caputa:2021ori, Caputa:2021sib,Balasubramanian:2022tpr, Muck:2022xfc, Adhikari:2022oxr, Adhikari:2022whf}.

Unitary evolution under a quantum Hamiltonian sends an initial operator ${\cal O}_0$ to its Heisenberg-evolved time-dependent version $e^{i Ht}{\cal O}_0 e^{-i Ht}$, exploring thus over time the space spanned by successive commutators of the form $\left[H, \left[ H,\cdots \left[ H,[ {\cal O}_0\right] \right] \right]$. In this way the initial operator ${\cal O}_0$ explores a larger and larger subspace of the Hilbert space of operators, and a natural notion of complexity should quantify how quickly this spread occurs, and furthermore how big a subspace of the Hilbert space of operators is eventually explored. As described in section \ref{sec.ReviewKcomplexity} below, K-complexity captures exactly this notion of spread mathematically and allows us to quantitatively distinguish different physical systems by the efficiency of this spread. This is done by transforming the intuitive idea of exploring higher and higher commutators, as above, into an orthogonal basis of the Hilbert space of operators, and studying the Heisenberg dynamics with respect to this basis.

In addition to the behavior of K-complexity for given individual quantum systems, such as the SYK model \cite{Parker:2018yvk, Jian:2020qpp, Rabinovici:2020ryf}, 2D CFTs \cite{PhysRevD.104.L081702, Caputa:2021ori}, and more general symmetry-based Hamiltonian systems \cite{Caputa:2021sib,Balasubramanian:2022tpr}, it is interesting and important to categorize the possible Krylov phenomenologies according to more universal criteria. One of the most interesting of these is clearly the behavior of K-complexity in the class of chaotic quantum systems as opposed to that of integrable ones, initiated in \cite{Rabinovici:2021qqt} for systems away from the thermodynamic limit. 

Quantum integrable systems, such as the strongly interacting XXZ chain \cite{Samaj_bajnok_2013} or the quadratic SYK model, are less efficient at exploring Krylov space, as evidenced for example by their reaching a lower saturation value of K-complexity at late times. By mapping the dynamics of operator spreading to an off-diagonal Anderson-like hopping problem on the Krylov chain this under-saturation is linked to the (partial)  localization of the wave function on the Krylov chain  \cite{Rabinovici:2021qqt}. Maximally chaotic systems feature a late-time complexity saturation value which is exponential in the number of degrees of freedom \cite{Rabinovici:2020ryf}; interacting integrable systems saturate at quantitatively lower values as compared to chaotic models due to localization effects in Krylov space \cite{Rabinovici:2021qqt}, and free systems typically depict complexity saturation values at late times which are linear, or polynomial, in the number of degrees of freedom \cite{Rabinovici:2020ryf}.  Related work, in the thermodynamic limit, includes \cite{Trigueros:2021rwj} for systems featuring many-body localization, and \cite{Noh_2021} for the transverse-field Ising model with and without an integrability breaking term. We will focus on K-complexity at long time scales for finite systems away from the thermodynamic limit \cite{Barbon:2019wsy, Rabinovici:2020ryf, Rabinovici:2021qqt}.

In this paper we explore K-complexity in a class of quantum systems that show integrable to chaotic phase transitions as a function of certain control parameters. We also characterise, for the purpose of comparison, the behavior of K-complexity in random matrix theory, both with and without time reversal symmetry. Interestingly we find agreement between the late-time behavior of a quantum chaotic Hamiltonian with that in the random matrix ensemble of the right symmetry class.

\begin{figure}
    \centering
    \includegraphics[scale=0.5]{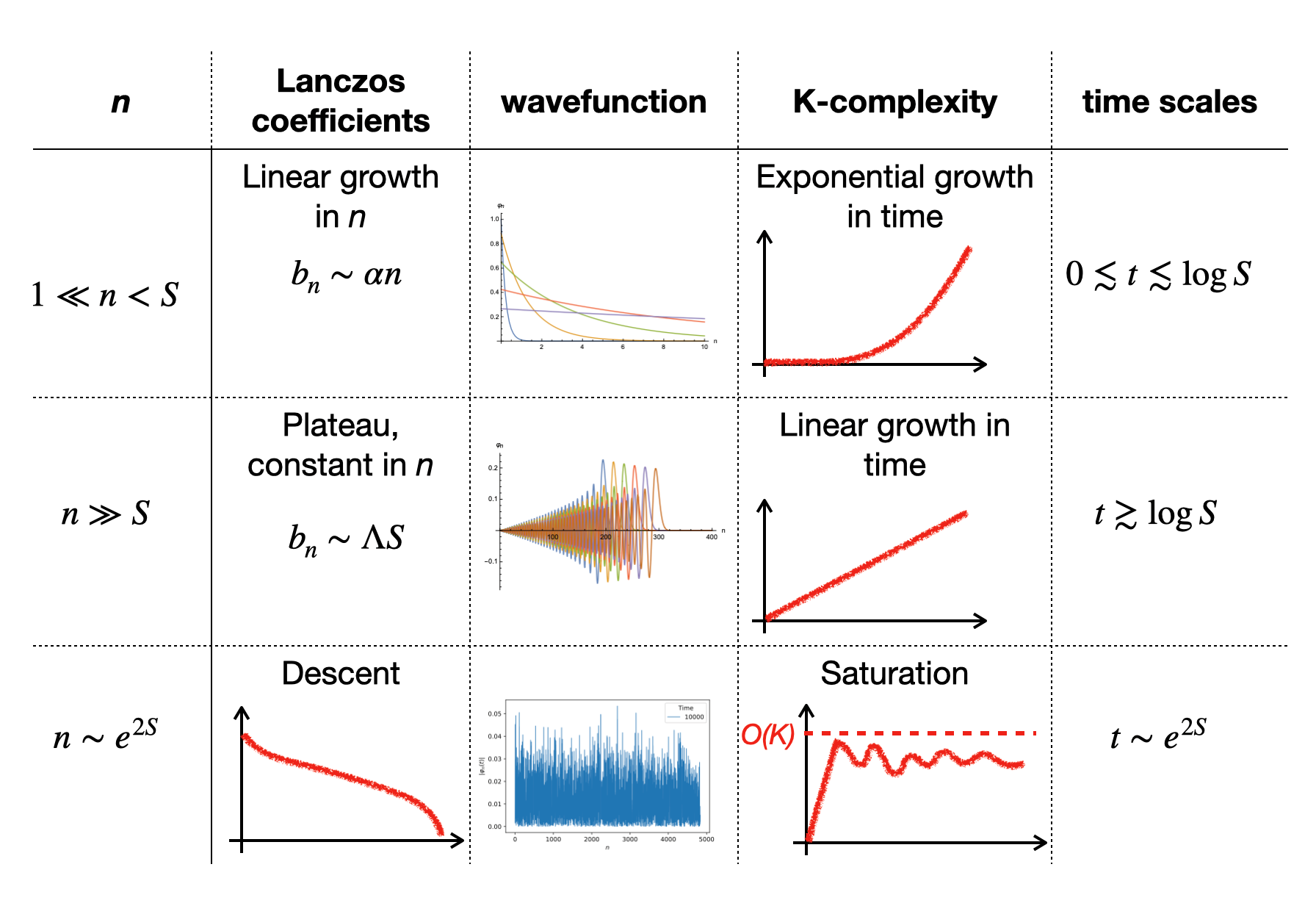}
    \caption{This table summarizes the general behaviour expected from K-complexity, particularly for finite chaotic systems with $S$ degrees of freedom ($\Lambda$ is the bandwidth of the system). It is based on \cite{Parker:2018yvk, Barbon:2019wsy} and \cite{Rabinovici:2020ryf}.}
    \label{fig:summary}
\end{figure}


In the remainder of this paper we will introduce and review background material regarding K-complexity (Section \ref{sec.ReviewKcomplexity}), and summarize what is known about the evolution as a function of time (see Figure \ref{fig:summary}). Section \ref{sec:XXZ_rstats} will introduce the main working horse of this study, the XXZ spin chain as well as two different integrability breaking deformations. In section \ref{sect_XXZ_KCsat} we will present numerical results performed on the integrable XXZ chain and its chaotic deformations exploring the behavior of K-complexity through the integrability-chaos transition with particular regard to the late-time saturation value. Section \ref{Sect:RMT} establishes the analogous results in pure random matrix theory (RMT), and categorizes the K-complexity behavior of chaotic systems with and without time reversal symmetry. We shall find agreement between the chaotic spin chain and the appropriate RMT universality class at sufficiently late time. We end with a discussion of our results in Section \ref{Sect:Disc}.

\section{Review of K-complexity and its late-time saturation value}\label{sec.ReviewKcomplexity}
Krylov complexity is a measure of operator complexification as it evolves in time. It is defined by constructing an orthonormal basis starting with the operator itself and constructing orthogonal directions by iteratively commuting it with the Hamiltonian.  It was introduced in \cite{Parker:2018yvk} as a probe of quantum chaos in the thermodynamic limit, and in \cite{Barbon:2019wsy} it was suggested as a measure of operator complexity at all time scales for finite systems with operators satisfying the Eigenstate Thermalization Hypothesis (ETH) \cite{Srednicki:1994mfb, Srednicki_1999, Peres_1984, Deutsch_1991, Sonner:2017hxc}.  In \cite{Rabinovici:2020ryf} K-complexity was computed for complex SYK$_4$ systems and it was shown numerically that
its time-dependent profile fits the one expected from quantum computation as well as from holography \cite{Susskind:2014rva, Susskind:2018pmk}. K-complexity was computed in \cite{Rabinovici:2021qqt} for the XXZ model which is a strongly interacting many-body integrable system, and was shown to saturate at late times at values below those found for SYK$_4$ which is a maximally chaotic system \cite{Kitaev_SYK, Sachdev_SYK, MaldacenaSYK}.  This paper aims to bridge the gap between the integrable and the chaotic by introducing a Hamiltonian which interpolates between the two, and studying K-complexity for a fixed type of local operator.

We now briefly review the definition of K-complexity. Given a Hamiltonian $H$, an operator $\mathcal{O}$ and an inner product $(\mathcal{A}| \mathcal{B}) = \frac{1}{D} \mathrm{Tr}(\mathcal{A}^\dagger \mathcal{B})$ where $D$ is the Hilbert space dimension, the Krylov basis is defined by an iterative orthonormalization procedure known as the Lanczos algorithm:
\begin{enumerate}
    \item $\mathcal{O}_0=\mathcal{O}/\|\mathcal{O}\|$
    \item For $n \geq 1$:
    $\mathcal{A}_n = [H, \mathcal{O}_{n-1}] - b_{n-1}\mathcal{O}_{n-2}$\\
    Compute $\|\mathcal{A}_n\|$\\
    If $\|\mathcal{A}_n\|=0$: STOP\\
    Otherwise: define norm $b_n=\|\mathcal{A}_n\|$ and normalized operator $\mathcal{O}_n = \mathcal{A}_n/b_n$.
\end{enumerate}
Here, $\|\mathcal{A}\| = \sqrt{(\mathcal{A} | \mathcal{A})}$ is the norm of an operator $\mathcal{A}$ and it is to be understood that $b_0=0$ and $\mathcal{O}_{-1}=0$. In this way we construct a complete \textit{ordered} orthonormal basis, the \textit{Krylov chain}, adapted to the operator's time-evolution. The value of $n$ at which the algorithm terminates is the \textit{Krylov space dimension}, denoted by $K$, and in \cite{Rabinovici:2020ryf} it was shown that it satisfies $K\leq D^2-D+1$. This bound is saturated in all the cases studied in this paper. The orthonormalization coefficients $b_n$ are called the \textit{Lanczos coefficients}.

The time-evolving operator can now be expanded in the Krylov basis 
\begin{equation} \label{time_evol}
    \mathcal{O}(t) = e^{iHt} \mathcal{O}_0 e^{-iHt}= \sum_{n=0}^{K-1} \phi_n(t)\mathcal{O}_n
\end{equation}
where $\phi_n(t)$ can be thought of as the wavefunction over the Krylov basis, which satisfies, via the Heisenberg equation, a Schr\"odinger-like equation
\begin{equation} \label{Phi_DE}
    -i\dot{\phi}_n(t) = b_n \phi_{n-1}(t) + b_{n+1} \phi_{n+1}(t)
\end{equation}
with boundary conditions $\phi_{-1}(t)=0$ and $\phi_{n}(t=0)=\delta_{0n}$.  From Unitarity, since the initial operator is normalized at the first step of the Lanczos algorithm, the wavefunction $\phi_n(t)$ is normalized at all times: $\sum_{n=0}^{K-1} |\phi_n(t)|^2=1$.

K-complexity is defined as the time-dependent average position over the Krylov chain
\begin{equation} \label{KC_def}
    C_K(t) = \sum_{n=0}^{K-1} n |\phi_n(t)|^2 ~.
\end{equation}
The behavior of $C_K(t)$ at different time scales for finite (chaotic) systems with $S$ degrees of freedom is summarized in Figure \ref{fig:summary}.  It is associated with the behavior of the Lanczos coefficients at different $n$ scales via the wavefunction (shown schematically in the same figure).

The Krylov elements $\{\mathcal{O}_n\}_{n=0}^{K-1}$ can be thought of as a basis of sites $|\mathcal{O}_n)$ in a chain of length $K$.  The action of the \textit{Liouvillian} $\mathcal{L} \equiv [H,\,\,]$ relates different sites on the Krylov chain i.e. $\mathcal{L}|\mathcal{O}_{n-1}) = b_n|\mathcal{O}_n) + b_{n-1}|\mathcal{O}_{n-2})$ and its matrix is tridiagonal. We shall denote the eigenvalues of the Liouvillian by $\omega_i$ and its eigenvectors by $|\omega_i)$,
\begin{equation}
    \mathcal{L}|\omega_i) = \omega_i|\omega_i) \quad i=0,\dots, K-1~.
\end{equation}
Note that the eigenvalues of the Liouvillian in Krylov space are equal to precisely those energy differences of the Hamiltonian, $E_a-E_b$, for which $O_{ab} \neq 0$, where $O_{ab}$ are the matrix elements of the operator $\mathcal{O}$ in the Hamiltonian's energy basis $\{|E_a\rangle\}_{a=1}^{D}$. That is,
\begin{eqnarray}
\label{Op_Specdec}
    \mathcal{O}&=&\sum_{a,b=1}^D  O_{ab}|E_a\rangle \langle E_b| ~.
\end{eqnarray}

In this formulation, the time-evolution of the operator is given by
\begin{equation}
    |\mathcal{O}(t))= e^{i\mathcal{L}t} |\mathcal{O}_0) = \sum_{i=0}^{K-1} e^{i\omega_i t}|\omega_i)(\omega_i|\mathcal{O}_0)~.
\end{equation}
From (\ref{time_evol}), the wavefunction $\phi_n(t)$ is the projection $(\mathcal{O}_n|\mathcal{O}(t))$. Hence the time-dependent transition amplitude is
\begin{equation}
    |\phi_n(t)|^2 = \sum_{i,j=0}^{K-1} e^{i(\omega_j-\omega_i)t} (\mathcal{O}_n|\omega_j)(\omega_j|\mathcal{O}_0) (\mathcal{O}_0|\omega_i)(\omega_i|\mathcal{O}_n)
\end{equation}
and the long-time average of $|\phi_n(t)|^2$ is given by
\begin{equation} \label{Q0n_LTA}
    Q_{0n}\equiv    \overline{|\phi_n|^2} = \lim_{T\to \infty} \frac{1}{T} \int_0^T |\phi_n(t)|^2 dt = \sum_{i=0}^{K-1}  |(\mathcal{O}_0|\omega_i)|^2 |(\omega_i|\mathcal{O}_n)|^2 ~.
\end{equation}
Note that the late-time average of the transition amplitude is normalized,
\begin{eqnarray}\label{Q0n_norm}
    \sum_{n=0}^{K-1}Q_{0n} = 1~.
\end{eqnarray}
From (\ref{KC_def}) and (\ref{Q0n_LTA}) the late-time saturation value of K-complexity is 
\begin{equation} \label{KC_LTA}
    \overline{C_K} = \sum_{n=0}^{K-1} n \overline{|\phi_n|^2} = \sum_{n=0}^{K-1} n Q_{0n} ~,
\end{equation}
which will be the main object of study of this paper. Before moving on to a more concrete study of K-complexity, we need to clarify the role of the connected and disconnected contributions to the various quantities just introduced.
\subsection{Effect of operator's trace on late-time saturation value of K-complexity}
In this section we discuss the effect of the operator's trace on the late-time saturation value of K-complexity.
It can be directly linked to the influence of the disconnected part of the two-point function on its late-time plateau and, just like when studying the latter, in order to probe universal effects due to chaotic behavior, one may work with operators with a zero one-point function or subtract it explicitly if it is initially non-zero.
For a hermitian normalized operator written in the energy basis as in (\ref{Op_Specdec}), the two-point function is given by
\begin{eqnarray}
\label{2Pt-function}
    \phi_0(t)&=&\frac{1}{D} \mathrm{Tr}\big[\mathcal{O}^\dagger \mathcal{O}(t) \big] = \frac{1}{D} \sum_{a,b=1}^D |O_{ab}|^2 e^{i(E_a-E_b)t} ~.
\end{eqnarray}
From which $Q_{00}$ is obtained by setting $n=0$ in (\ref{Q0n_LTA}):
\begin{eqnarray}
\label{Q00_LTA}
    Q_{00}&\equiv& \overline{|\phi_0|^2} = \lim_{T\to \infty}\frac{1}{T} \int_0^T |\phi_0(t)|^2 dt \nonumber\\
    &=&\frac{1}{D^2} \lim_{T\to \infty}\frac{1}{T} \int_0^T  \sum_{a,b,c,d=1}^D |O_{ab}|^2 |O_{cd}|^2 e^{i(E_a-E_b+E_c-E_d)t} dt \nonumber\\
    &=& \frac{1}{D^2} \Big[ \sum_{a,b=1}^D  |O_{aa}|^2 |O_{bb}|^2 + \sum_{a\neq b=1}^D  |O_{ab}|^4  \Big]~,
\end{eqnarray}
where in the last step we assumed the absence of degeneracies or rational relations in the energy spectrum.
In Appendix \ref{appx_Connected} it is shown that using the ``ETH" ansatz for RMT \cite{DAlessio:2015qtq}, i.e.
\begin{equation}
    \centering
    \label{Op_RMT_Ansatz}
    O_{ab} = O\delta_{ab}+\frac{1}{\sqrt{D}}r_{ab}~,
\end{equation}
where $O$ is $O(1)$ and the matrix $r_{ab}$ is drawn from a Gaussian ensemble with zero mean and unit variance, the scaling of $Q_{00}$ is as follows
\begin{eqnarray} \label{Q00_ETH}
     Q_{00} &\sim& O(1)+ O\left(\frac{1}{D}\right)~.
\end{eqnarray}

In the case of the two-point function, to better probe the spectral correlations in the system, one usually studies the \textit{connected} part which amounts to using the traceless operator:
\begin{equation} \label{Op_TL}
    \tilde{\mathcal{O}} \equiv \mathcal{O} - \frac{1}{D} \textrm{Tr}(\mathcal{O})\mathbb{1}~.
\end{equation}

Using the traceless version of the operator (\ref{Op_TL}), it is shown in Appendix \ref{appx_Connected} that together with the ansatz (\ref{Op_RMT_Ansatz}), the connected version of $Q_{00}$ behaves as
\begin{equation}
    \centering
    \label{Q00_Conn}
    Q_{00}^{(c)}\sim O\left(\frac{1}{D^2}\right)~.
\end{equation}
From (\ref{KC_LTA}), if  $Q_{00}$ is significantly large compared to $Q_{0n}$ for $n>0$ (taking (\ref{Q0n_norm}) into account), the saturation value of K-complexity will be pulled down to smaller values. From (\ref{Q00_ETH}) it is clear that $Q_{00}$ can be as large as permitted by normalization for an operator with non-zero trace, its value being controlled by the one-point function, but this does not reflect any universal behaviour of the autocorrelation function. In order to study universal features, one must work with operators with a zero one-point function, since in that case the two-point function is directly equal to its connected part, and ETH predicts that the latter plateaus at $\frac{1}{D}$, while the corresponding transition probability plateaus at $Q_{00}^{(c)}\sim\frac{1}{D^2}$. In chaotic systems, this is consistent with the observation that $Q_{0n}^{(c)}$ approaches $O\left(\frac{1}{D^2}\right) \sim\frac{1}{K}$ for all $n=0,...,K-1$.  This means that for chaotic systems the late-time transition probability is more uniform as a function of $n$, compatible with the normalization $\sum_{n=0}^{K-1}Q_{0n}=1$ and implying a K-complexity long-time average which approaches $\sim \frac{K}{2}$ as seen for example in complex SYK$_4$ \cite{Rabinovici:2020ryf}. For the effect of the 1-point function on the saturation value of K-complexity in the complex SYK$_4$ model see Appendix \ref{app:SYK}. 

\section{XXZ and its integrability breaking} \label{sec:XXZ_rstats}
The Heisenberg XXZ spin chain is an integrable model which exhibits Poisson level-spacing statistics.  The model consists of nearest-neighbor spin interactions
\begin{equation} \label{HXXZ}
    H_{XXZ} =  \sum_{i=1}^{N-1} J \left( S_i^x S_{i+1}^x + S_i^y S_{i+1}^y \right) + J_{zz} S_i^z S_{i+1}^z
\end{equation}
where $S_i^\alpha =1/2 \sigma_i^\alpha$ and $ \sigma_i^\alpha$ are the Pauli matrices with $\alpha=x,y,z$. 
In a series of papers it was shown that even the addition of a local operator such as
\begin{equation} \label{IBT_local}
    H_d =  S^z_{j}
\end{equation}
to the XXZ Hamiltonian can break its integrability \cite{Santos_2004, PhysRevE.84.016206, PhysRevB.80.125118, PhysRevB.98.235128, PhysRevB.102.075127, PhysRevX.10.041017} and spectral statistics will show chaotic behaviour. Another type of integrability breaking term \cite{doi:10.1119/1.3671068} we will consider is the next-to-nearest-neighbour operator
\begin{equation} \label{IBT_NL}
    H_{NNN} = \sum_{i=1}^{N-2} S_i^z S_{i+2}^z ~.
\end{equation}

We will demonstrate the transition from integrability to chaos by studying the distribution of the ratios of consecutive level spacings \cite{PhysRevLett.110.084101, PhysRevB.75.155111}, and show that increasing the strength of the integrability breaking term from zero will result in a transition in the spectral behaviour from integrable to chaotic. For an ordered set of energy eigenvalues $\{E_i\}_{i=1}^D$, consecutive level spacings are defined as $s_i=E_{i+1}-E_i$ and consecutive ratios are defined as the set $r_i = s_{i}/s_{i-1}$. 
The distribution $P(r)$ was computed in \cite{PhysRevLett.110.084101} for the random matrix ensembles GOE, GUE and GSE.
It is useful to define the quantity $\tilde{r_i} = \min\left( r_i, \frac{1}{r_i}\right)$ with distribution $P(\tilde{r})=2P(r)\theta(r-1)$ whose mean $\langle \tilde{r} \rangle$ can be used as an indicator to distinguish an integrable system from a chaotic one. For a Poissonian distribution of level-spacings $P(s)= e^{-s}$, the distribution of $r$ is given by $P(r)= (1+r)^{-2}$ \cite{PhysRevB.75.155111} and $\langle \tilde{r}\rangle = 2 \ln 2 -1 \approx 0.38629$.  For the Wigner ensembles (GOE, GUE and GSE) distinguished by their Dyson index ($\beta= 1,2$ and $4$ respectively) it was shown in \cite{PhysRevLett.110.084101} that a very good approximation for practical purposes is $P(r)=\frac{1}{Z_\beta} \frac{(r+r^2)^\beta}{(1+r+r^2)^{1+\frac{3}{2}\beta}}$, where $Z_\beta$ is a normalization constant. The $\langle \tilde{r}\rangle$ value for GOE is approximately $0.53590$.

\subsection{Choice of sector and local operator}
The XXZ Hamiltonian commutes with the operator representing the total spin in the $z$-direction 
\begin{equation}
    M = \sum_{i=1}^N S_i^z
\end{equation}
and is invariant under reflection with respect to the edge of the chain, represented by the parity operator $P$ \cite{doi:10.1119/1.4798343}.  To avoid degeneracies in the Hamiltonian spectrum we will work in a sector with fixed total spin and parity.  To study K-complexity we will use open boundary conditions and focus on a local operator $\mathcal{O}$ which respects these two symmetries and keeps the computation within the chosen sector:
\begin{equation} \label{Operator}
    \mathcal{O} = S_i^z + S_{N-i+1}^z~,
\end{equation}
where $i$ is chosen to be near the center of the chain.  This operator has non-zero trace and we remove its trace according to (\ref{Op_TL}) before performing the Lanczos algorithm. Appendix \ref{app:XXZ} discusses the effect of the one-point function on the saturation value of K-complexity in pure XXZ.

When adding the integrability breaking term (\ref{IBT_local}), we keep within the sector by using an odd-length chain and situating the impurity at the middle of the chain: 
\begin{equation} \label{IBT_Lmid}
    H_d = S_{(N+1)/2}^z ~.
\end{equation}
The integrability breaking term (\ref{IBT_NL}) commutes both with $M$ and with $P$. 

\subsection{r-statistics for XXZ with integrability breaking terms}
We now present results for the r-statistics of the following interpolating Hamiltonians:
\begin{equation} \label{XXZ+Hd}
    H = H_{XXZ} +\epsilon_d H_d
\end{equation}
and
\begin{equation} \label{XXZ+NNN}
    H = H_{XXZ} + J_{zz}^{(2)} H_{NNN}
\end{equation}
where $H_{XXZ}$ is given in (\ref{HXXZ}), $H_d$ in (\ref{IBT_Lmid}) and $H_{NNN}$ is given in (\ref{IBT_NL}).  We work with various values of $J_{zz}$ and set $J=1$ in all cases. Some of the Hamiltonians and operators used in the numerical computations were constructed using the QuSpin package \cite{QuSpin}. The Lanczos algorithm and K-complexity computations were performed using the codes we developed in \cite{Rabinovici:2020ryf} and \cite{Rabinovici:2021qqt}.

\begin{figure}
    \centering
    \begin{subfigure}[t]{\textwidth}
    \centering
        \includegraphics[scale=0.8]{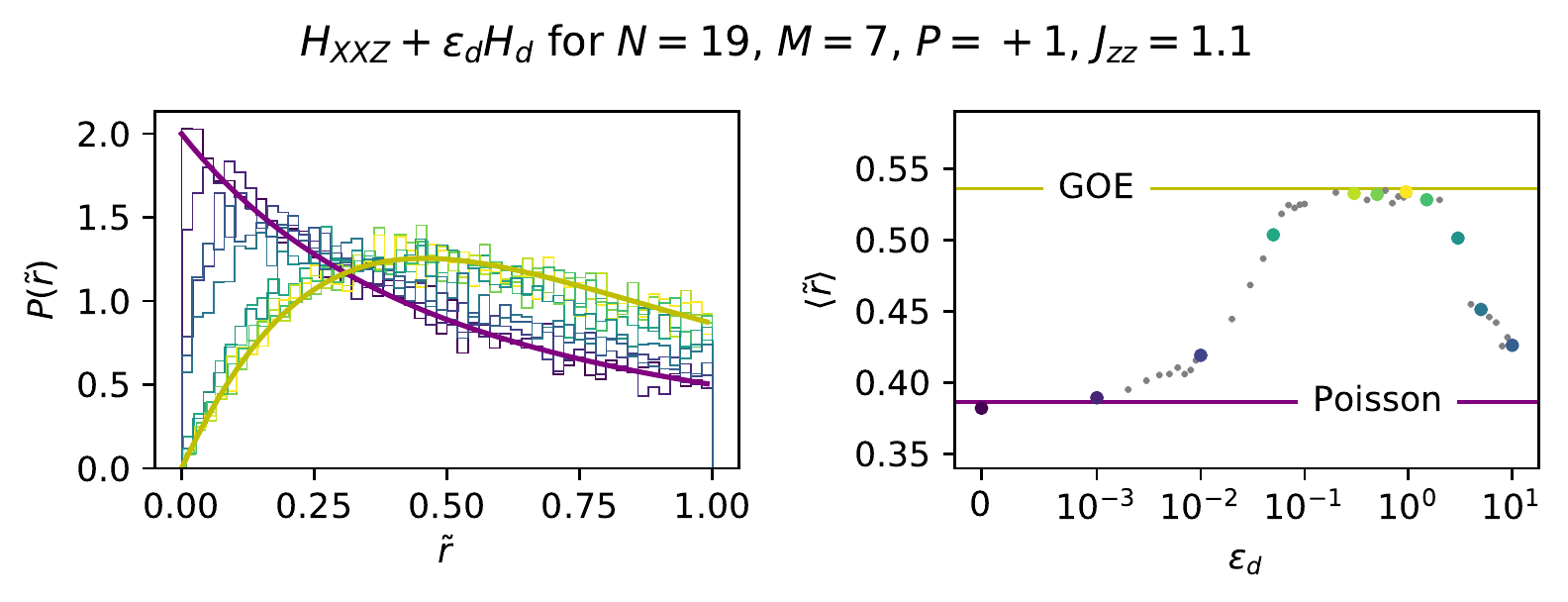}
        \caption{ $\tilde{r}$ statistics for $H_{XXZ}+\epsilon_d H_d$.}
        \label{fig:rStats_Hd}
    \end{subfigure}
    \begin{subfigure}[t]{\textwidth}
        \centering
    \includegraphics[scale=0.8]{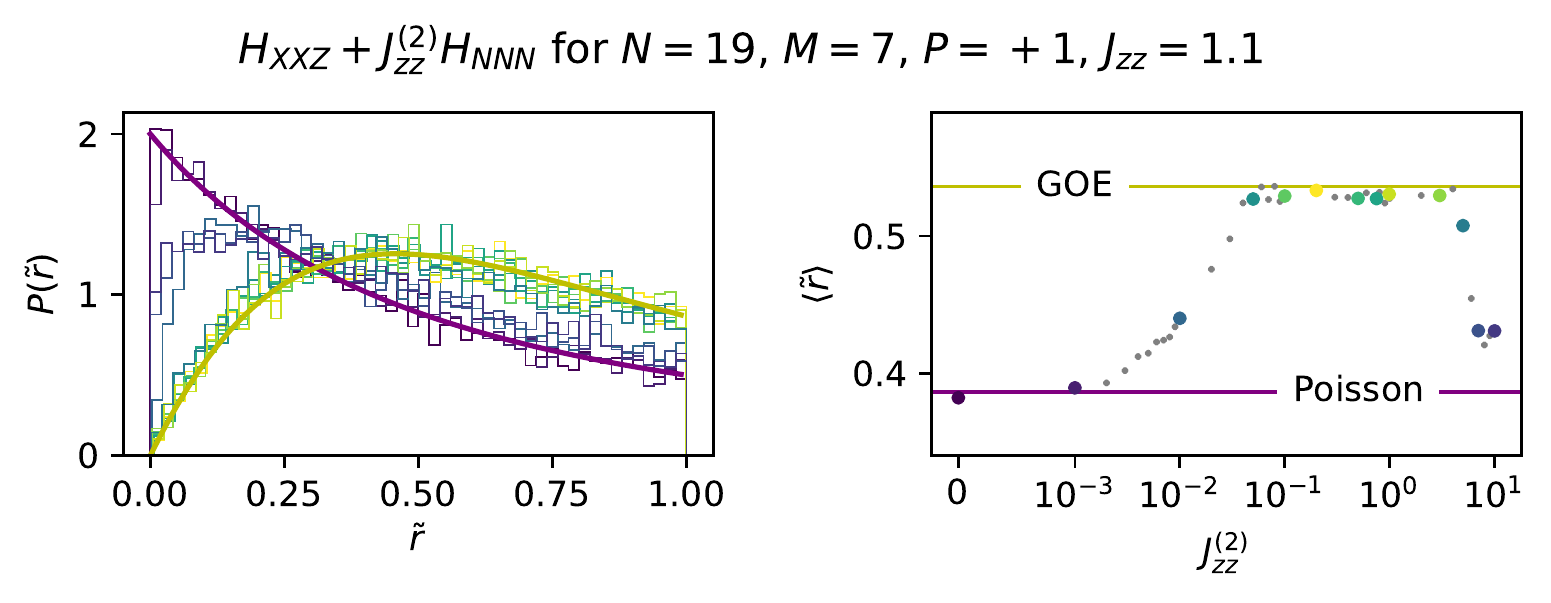}
    \caption{ $\tilde{r}$ statistics for $H_{XXZ}+J^{(2)}_{zz}H_{NNN}$.}
    \label{fig:rStats_NNN}
    \end{subfigure}
    \caption{ \textbf{Left:} Probability distribution functions for the $\tilde{r}$ statistics of  (\ref{XXZ+Hd}) (top) and (\ref{XXZ+NNN}) (bottom) with increasing value of $\epsilon_d$ and $J^{(2)}_{zz}$ respectively, computed for $N=19$ spins in the $M=7, P=+1$ sector with $J_{zz}=1.1$. The purple line represents the analytical result for $P(\tilde{r})$ in the case of Poissonian level-spacing statistics, while the yellow line represents the analytical result for GOE ensembles. \textbf{Right:} The value of $\langle\tilde{r}\rangle$ as a function of $\epsilon_d$ (top) and $J^{(2)}_{zz}$ (bottom). Horizontal lines represent analytical values for Poisson (purple) and GOE (yellow). The colored dots represent point for which we plotted the $P(\tilde{r})$ distribution function in the left panel, while the gray dots represent additional data points.}
    \label{fig:rStats}
\end{figure}

Figures \ref{fig:rStats_Hd} and \ref{fig:rStats_NNN} show the $\tilde{r}$ statistics for the Hamiltonians (\ref{XXZ+Hd}) and (\ref{XXZ+NNN}) respectively.  We plot the distributions of $\tilde{r}$ for various values of the coefficient of the integrability breaking terms, as well as the mean values $\langle\tilde{r}\rangle$. We compare the results for both $P(\tilde{r})$ and $\langle\tilde{r}\rangle$ with the analytical results for Poisson and GOE mentioned in Section \ref{sec:XXZ_rstats}.  We see that increasing the strength of the integrability breaking term makes the system transition from displaying integrable statistics to displaying chaotic statistics. Note that after the transition, increasing the value of the coefficient of the integrability breaking term even further makes the system less chaotic, as can be seen in Fig. \ref{fig:rStats}.

\section{K-complexity and integrability-chaos transition}\label{sect_XXZ_KCsat}
In \cite{Rabinovici:2021qqt} it was shown that the saturation value of K-complexity is sensitive to the integrability/chaos of a model, by comparing results for complex SYK$_4$ systems with results for XXZ systems of similar Krylov space dimensions.
It was argued that the time evolution on the Krylov chain given by Equation (\ref{Phi_DE}) can be mapped to an Anderson problem with off-diagonal disorder. Higher disorder would imply some amount of localization for the Liouvillian eigenvectors and hence a smaller saturation value of K-complexity, while less disorder would imply less localization and higher saturation values of K-complexity.  
In this section we study the Lanczos coefficient statistics and saturation value of K-complexity for the interpolating Hamiltonians given by (\ref{XXZ+Hd}) and (\ref{XXZ+NNN}), with an operator of the type (\ref{Operator}). By increasing the value of the coefficient of the integrability breaking term we interpolate from a fully integrable model (XXZ) to a chaotic model, as can be seen through the $\langle \tilde{r}\rangle$ transition in Figure \ref{fig:rStats}.   In Figure \ref{fig:Lanczos_stats} we plot the distribution of the log of ratios of consecutive Lanczos coefficients $\log(b_{n}/b_{n+1})$.  The mean of this distribution is $\approx 0$ and the standard deviation generally decreases with the strength of integrability breaking, indicating less disorder in the Lanczos sequence.  
\begin{figure}
    \centering
    \begin{subfigure}[t]{0.45\textwidth}
    \centering
        \includegraphics[scale=0.5]{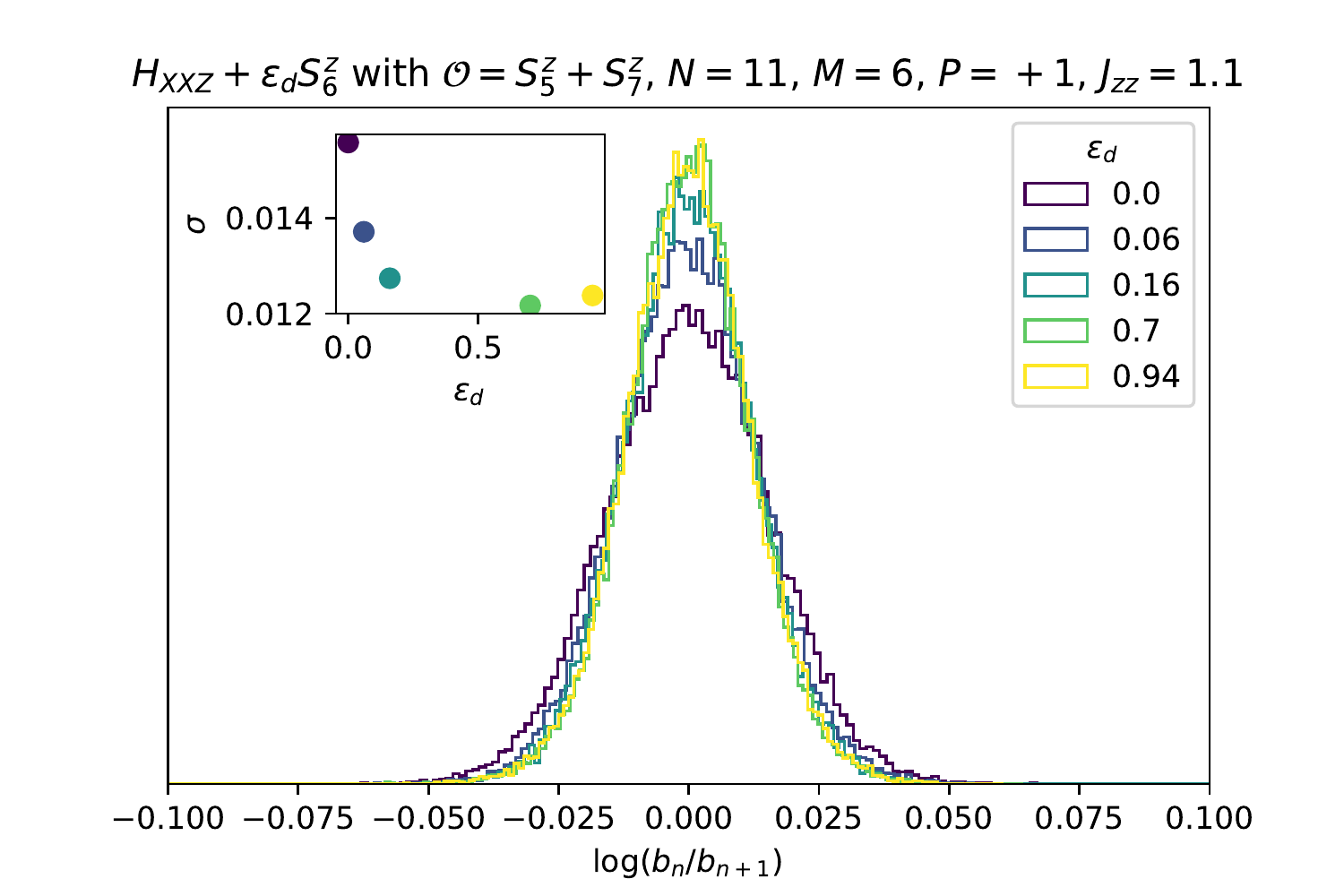}
        \caption{$H_{XXZ} +\epsilon_d H_d$ with $J_{zz}=1.1$ for $N=11$ spins in the $M=6, P=+1$ sector for the operator $\mathcal{O}=S_5^z+S_7^z$. }
    \label{fig:Lanczos_stats}
    \end{subfigure}
    \hfill
    \begin{subfigure}[t]{0.45\textwidth}
    \centering
        \includegraphics[scale=0.5]{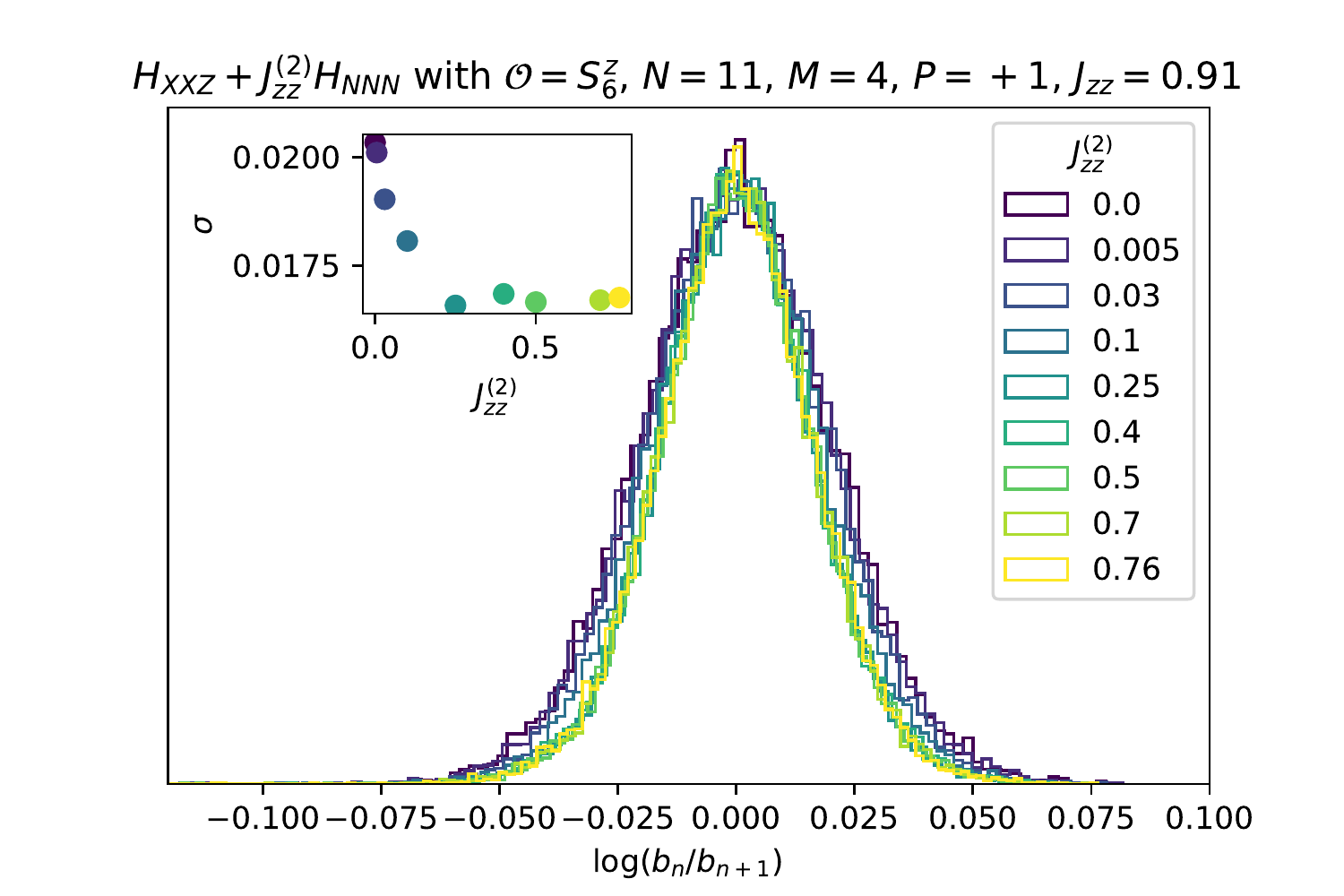}
        \caption{$H_{XXZ} +J_{zz}^{(2)} H_{NNN}$ with $J_{zz}=0.91$ for $N=11$ spins in the $M=4, P=+1$ sector for the operator $\mathcal{O}=S_6^z$.  }
    \label{}
    \end{subfigure}
    \caption{Distribution of the log of consecutive ratios of Lanczos coefficients. \textbf{Inset:} Standard-deviation $\sigma$ of this distribution as a function of the corresponding integrability breaking term.  The standard deviation generally decreases with the coefficient of the integrability breaking term.  Comparing with the corresponding computations of the K-complexity saturation values in Figures \ref{fig:N11M6Hd} and \ref{fig:N11M4J2zz}, this is consistent with the phenomenology described in \cite{Rabinovici:2021qqt} namely that the saturation values of K-complexity will increase with decreasing disorder in the Lanczos coefficients.}
    \label{fig:Lanczos_stats}
\end{figure}

\begin{figure}[t]
    \centering
    \begin{subfigure}[t]{0.45\textwidth}
    \centering
        \includegraphics[scale=0.5]{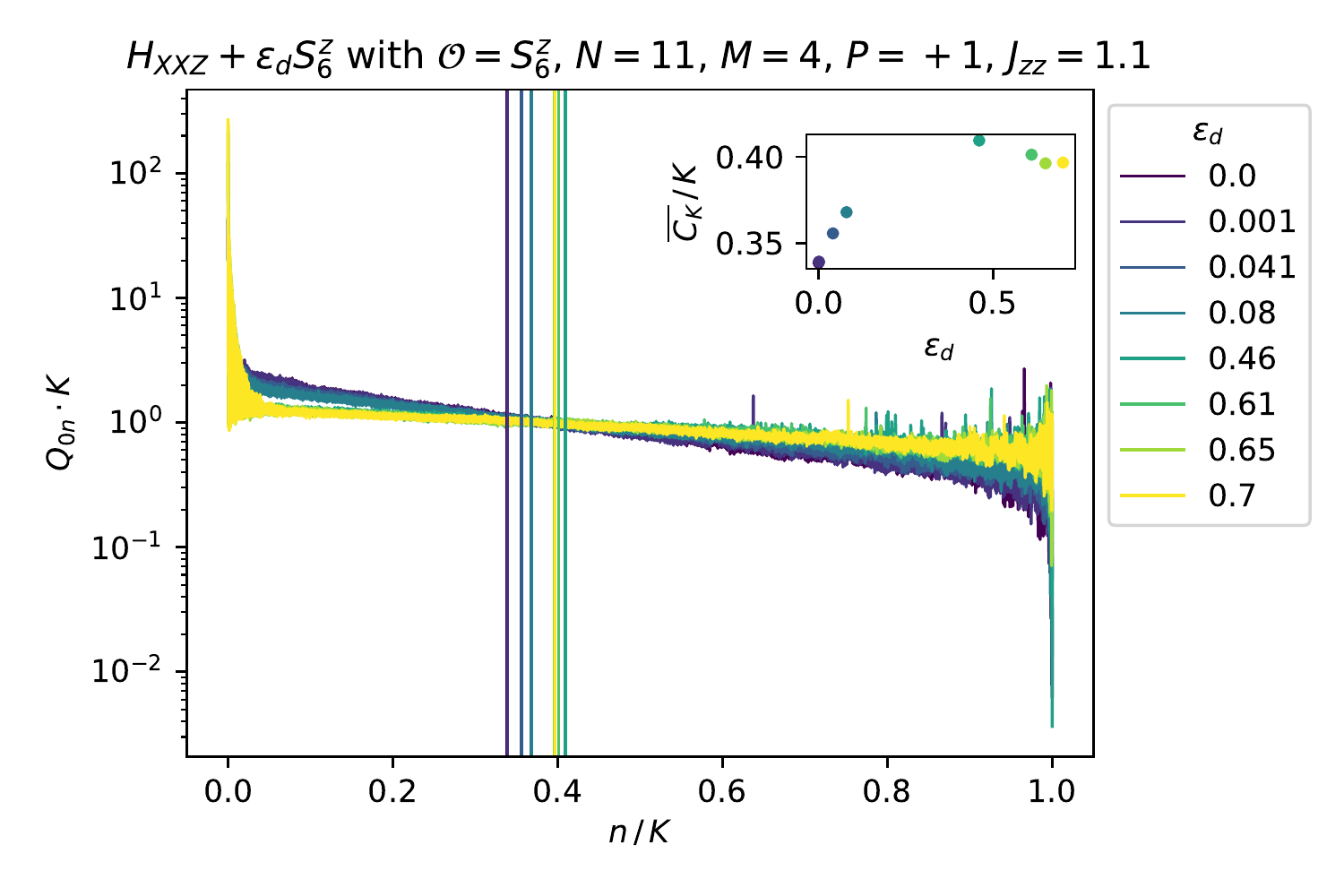}
        \caption{}
        \label{fig:N11M4Hd}
    \end{subfigure}
    \hfill
    \begin{subfigure}[t]{0.45\textwidth}
    \centering
        \includegraphics[scale=0.5]{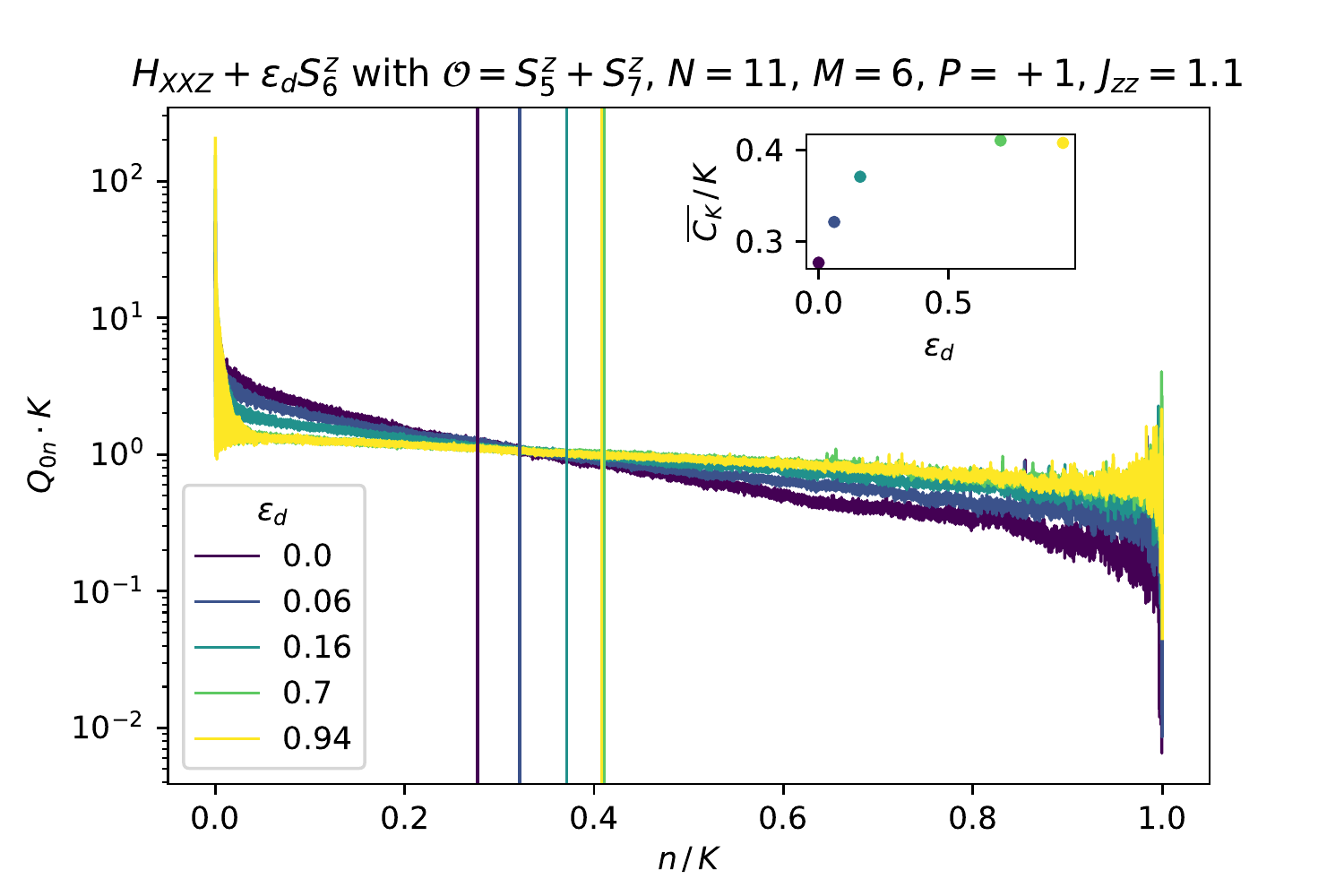}
        \caption{}
        \label{fig:N11M6Hd}
    \end{subfigure}
    \caption{Late-time transition probability results for local operator of the form (\ref{Operator}) with trace removed, for $H_{XXZ}$ with an $H_d$ integrability-breaking term. The vertical lines represent the late-time saturation value of KC as a fraction of the Krylov space dimension. \textbf{Left:} For $N=11$ spins in the sector $M=4, P=+1$ with $J_{zz}=1.1$ for the operator $\mathcal{O}=S_6^z$. The Krylov space dimension is $K=28731$. \textbf{Right:} For $N=11$ spins in the sector $M=6, P=+1$ with $J_{zz}=1.1$ for the operator $\mathcal{O}=S_5^z+S_7^z$. For this system the Krylov space dimension is $K=55461$. \textbf{Inset:} dependence of KC saturation value on the strength of the integrability-breaking term.}
    \label{fig:KC_sat_XXZ_Hd}
\end{figure}

\begin{figure}
    \centering
    \begin{subfigure}[t]{0.45\textwidth}
    \centering
        \includegraphics[scale=0.5]{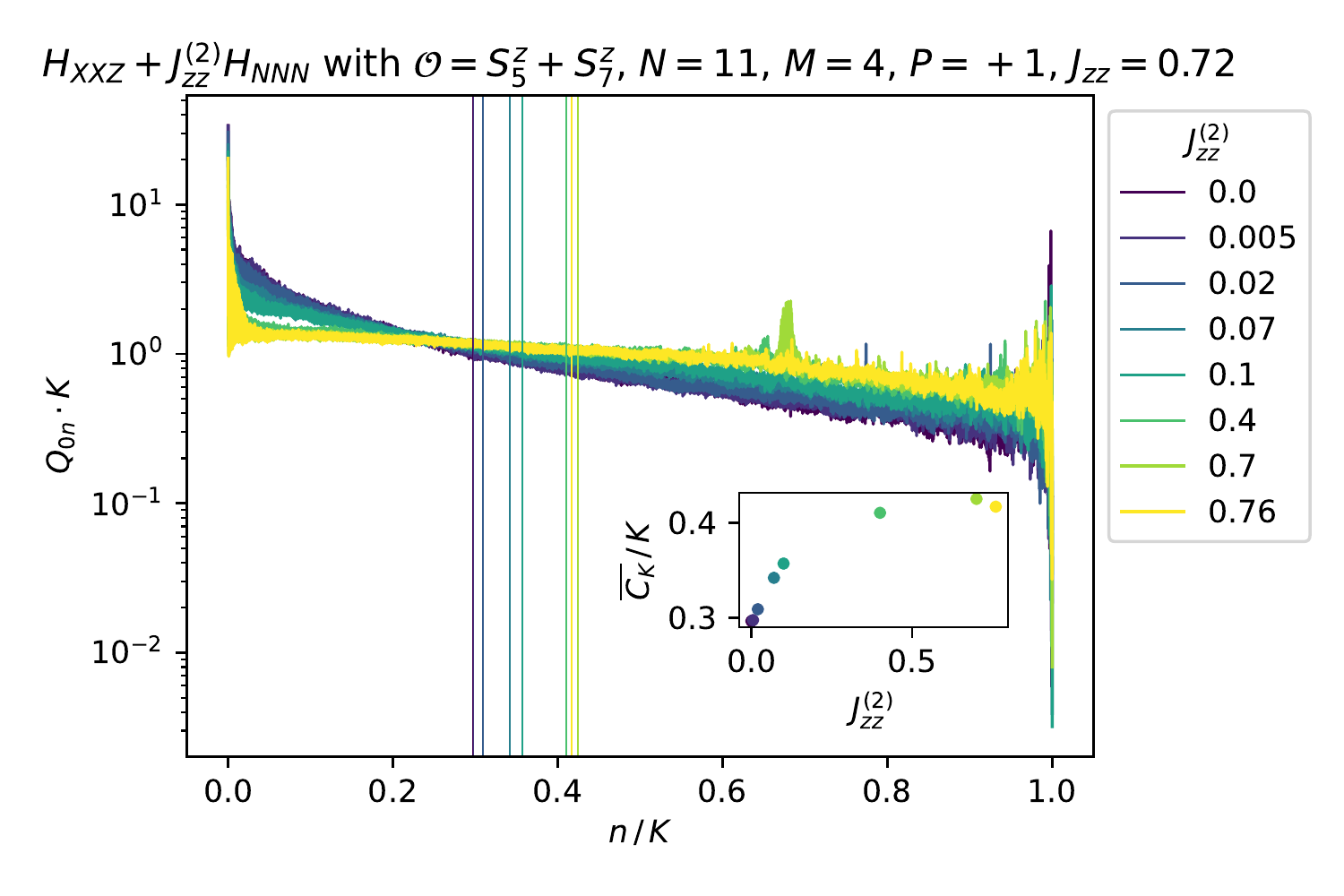}
        \caption{}
        \label{fig:}
    \end{subfigure}
    \hfill
    \begin{subfigure}[t]{0.45\textwidth}
    \centering
        \includegraphics[scale=0.5]{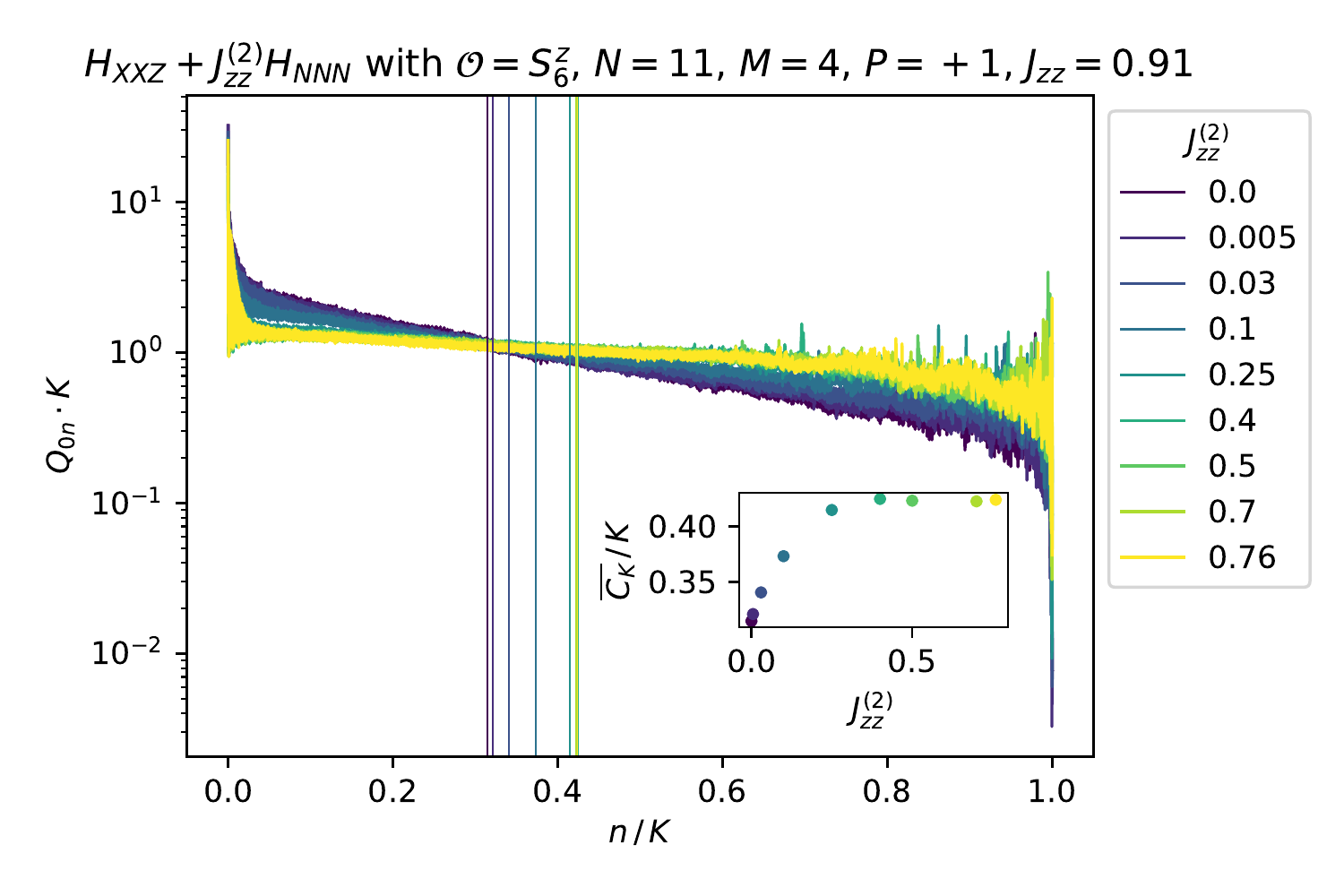}
        \caption{}
        \label{fig:N11M4J2zz}
    \end{subfigure}
    \caption{Results for the saturation value of K-complexity computed for a local operator of the form (\ref{Operator}) with trace removed, for $H_{XXZ}+J^{(2)}_{zz}\,H_{NNN}$ integrability-breaking term in the sector $N=11, M=4, P=+1$. \textbf{Left:} With $J_{zz}=0.72$, for the operator $\mathcal{O}=S_5^z+S_7^z$.   \textbf{Right:} With $J_{zz}=0.91$, for the operator $\mathcal{O}=S_6^z$. For both systems the Krylov space dimension is $K=28731$.
    \textbf{Inset:} dependence of KC saturation value on the strength of the integrability-breaking term.}
    \label{fig:KC_sat_XXZ_NNN}
\end{figure}

\begin{figure}
    \centering
    \centering
    \begin{subfigure}[t]{0.3\textwidth}
    \centering
        \includegraphics[scale=0.35]{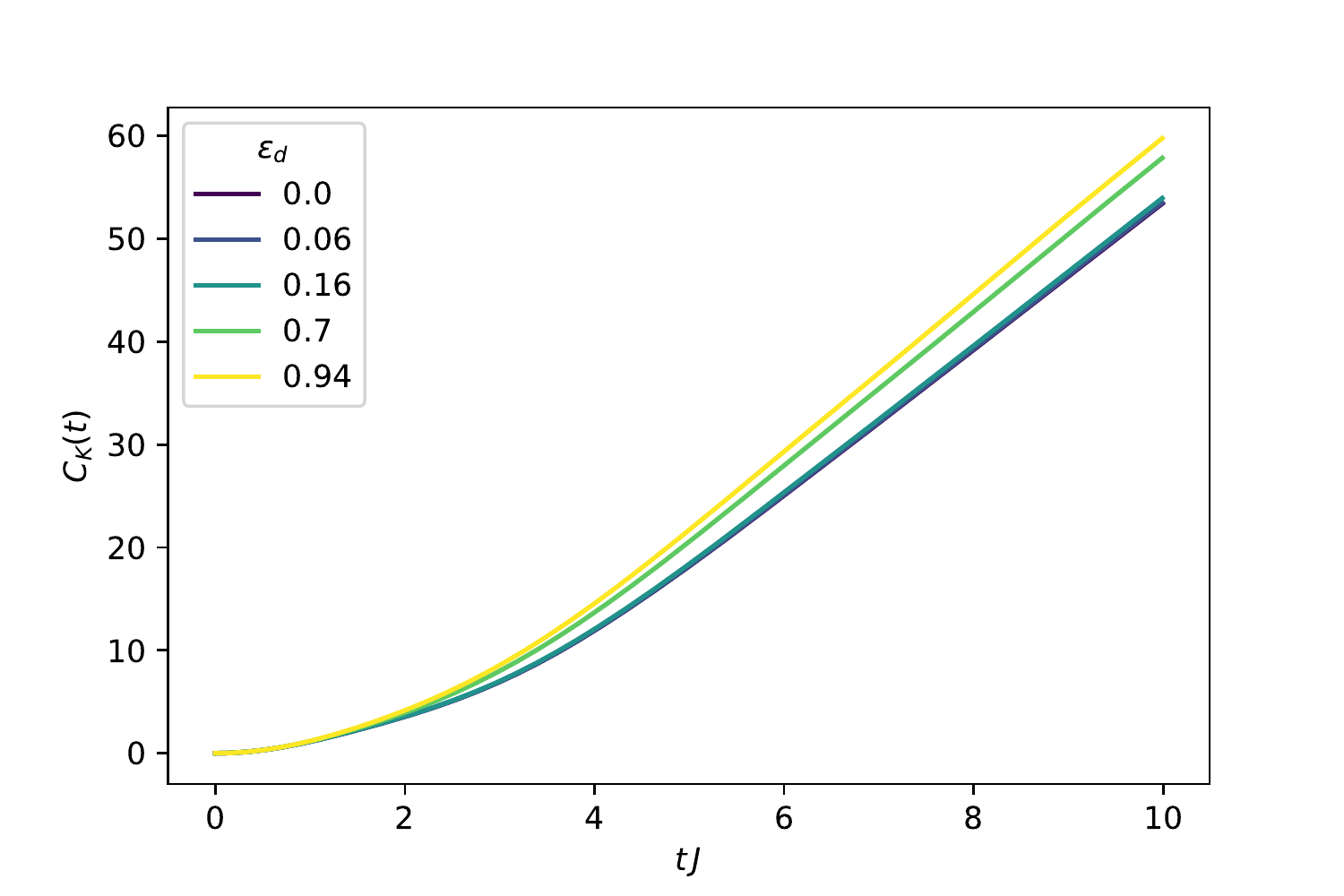}
        \caption{}
        \label{fig:}
    \end{subfigure}
    \hfill
    \begin{subfigure}[t]{0.3\textwidth}
    \centering
        \includegraphics[scale=0.35]{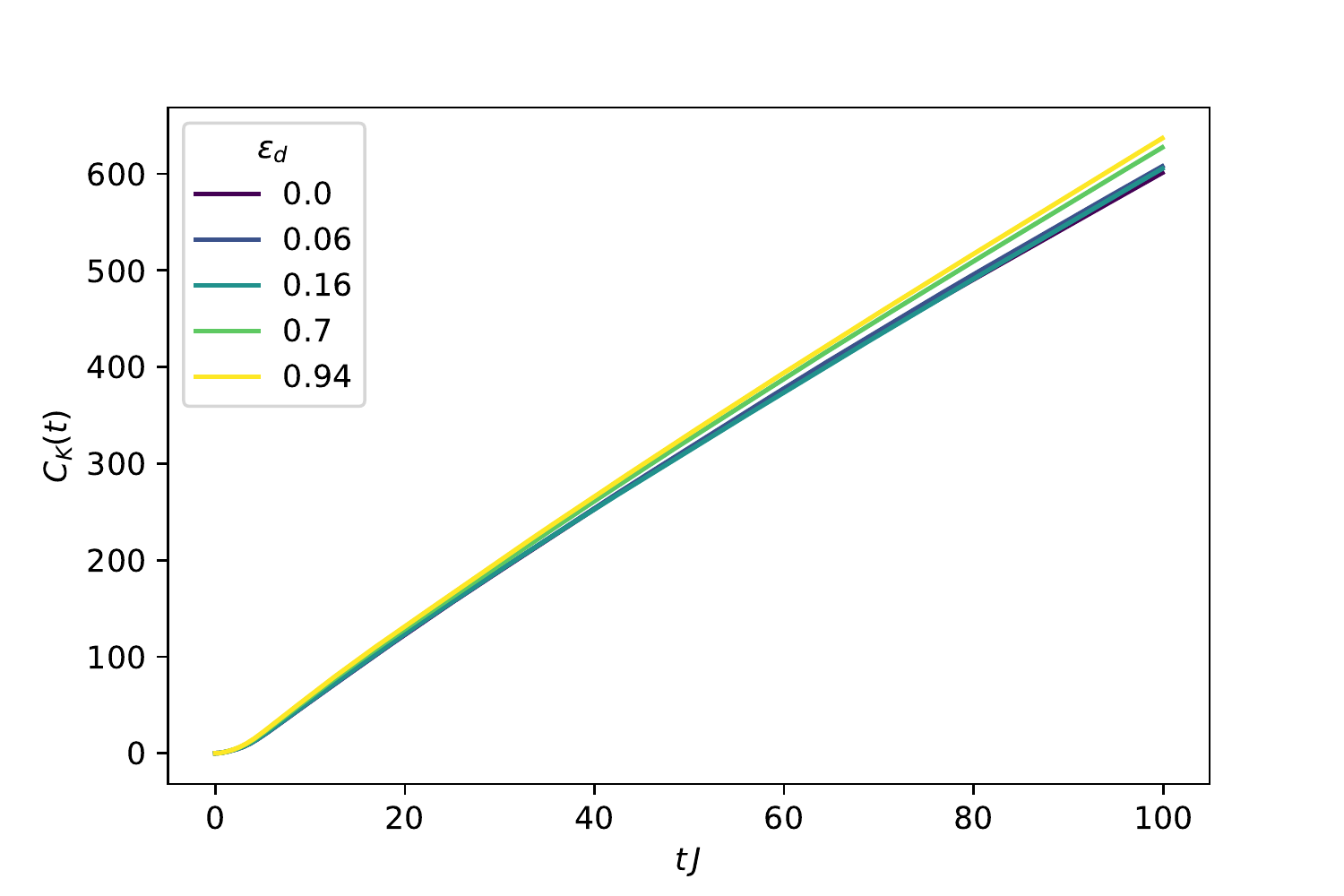}
        \caption{}
        \label{fig:}
    \end{subfigure}
    \hfill
    \begin{subfigure}[t]{0.3\textwidth}
    \centering
        \includegraphics[scale=0.35]{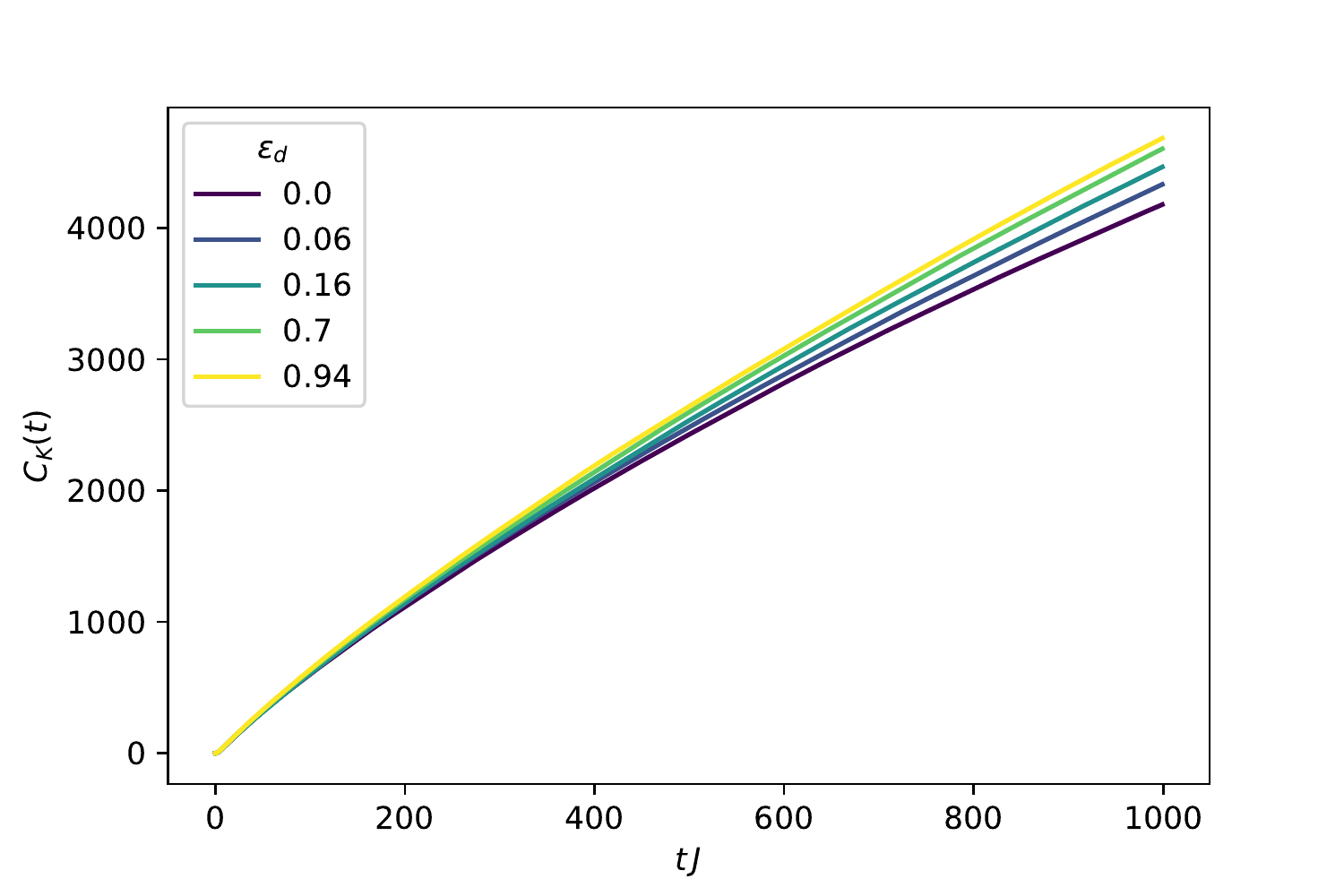}
        \caption{}
        \label{fig:}
    \end{subfigure}
    \vfill
    \begin{subfigure}[t]{0.3\textwidth}
    \centering
        \includegraphics[scale=0.35]{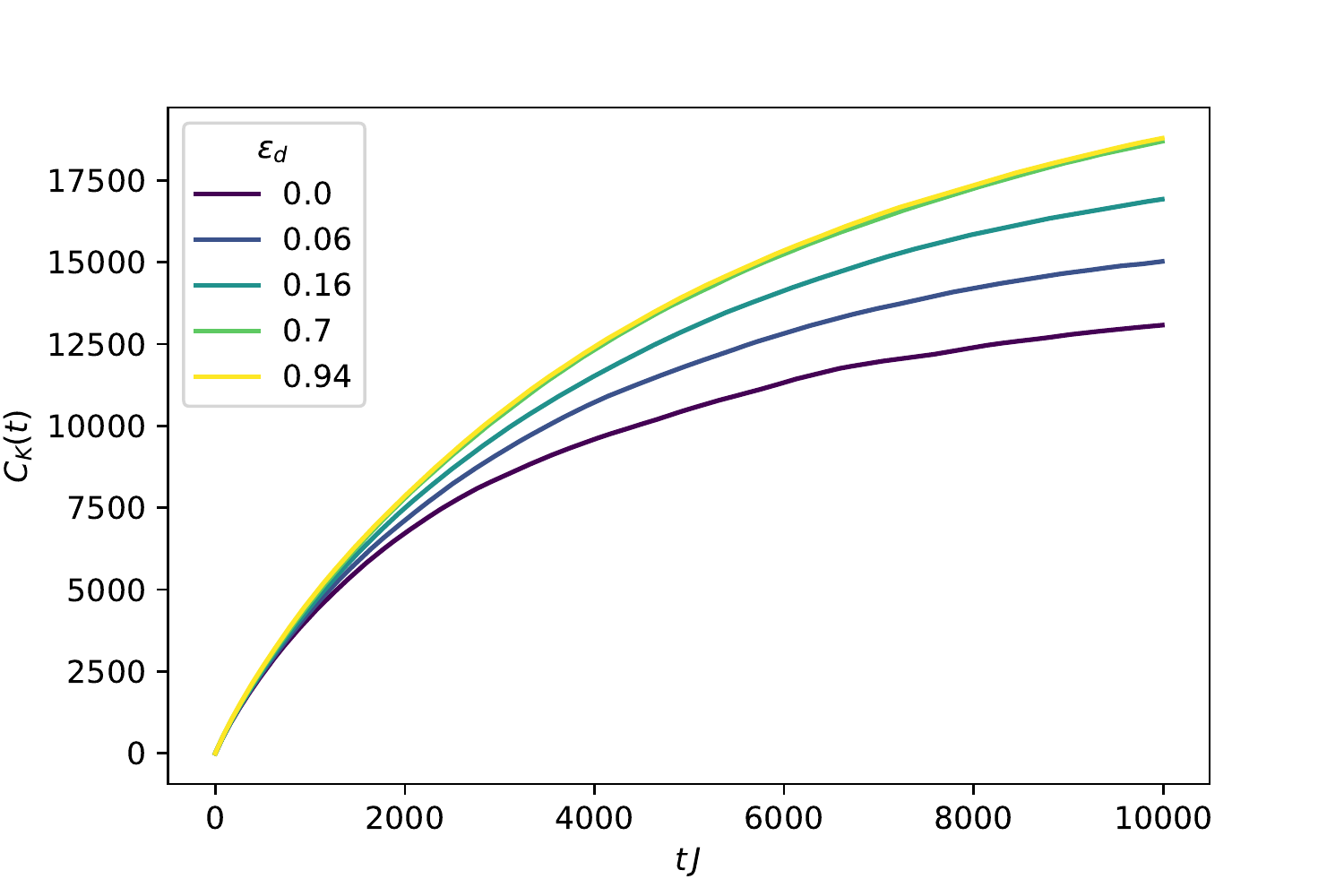}
        \caption{}
        \label{fig:}
    \end{subfigure}
    \hfill
    \begin{subfigure}[t]{0.3\textwidth}
    \centering
        \includegraphics[scale=0.35]{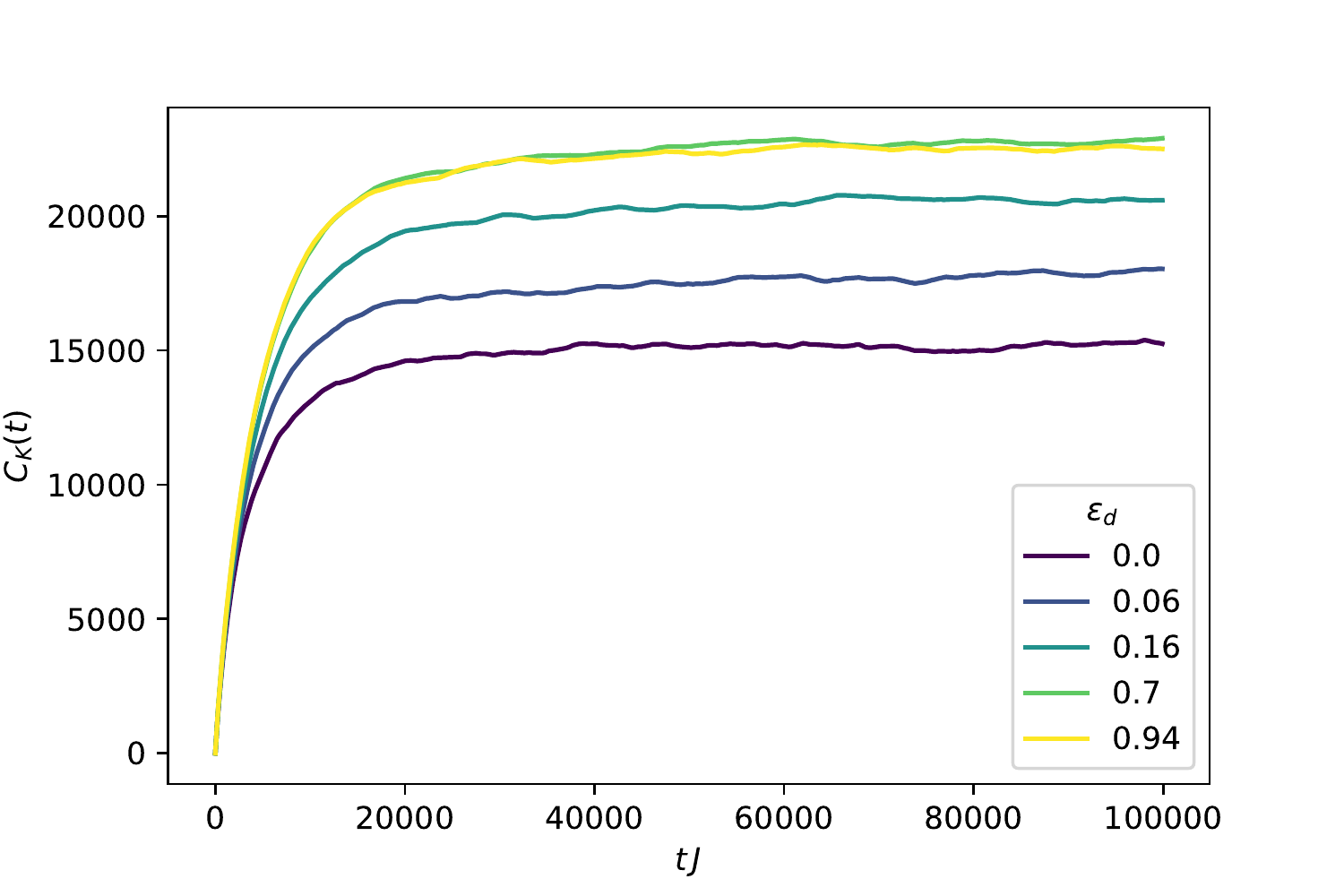}
        \caption{}
        \label{fig:}
    \end{subfigure}
    \hfill
    \begin{subfigure}[t]{0.3\textwidth}
    \centering
        \includegraphics[scale=0.35]{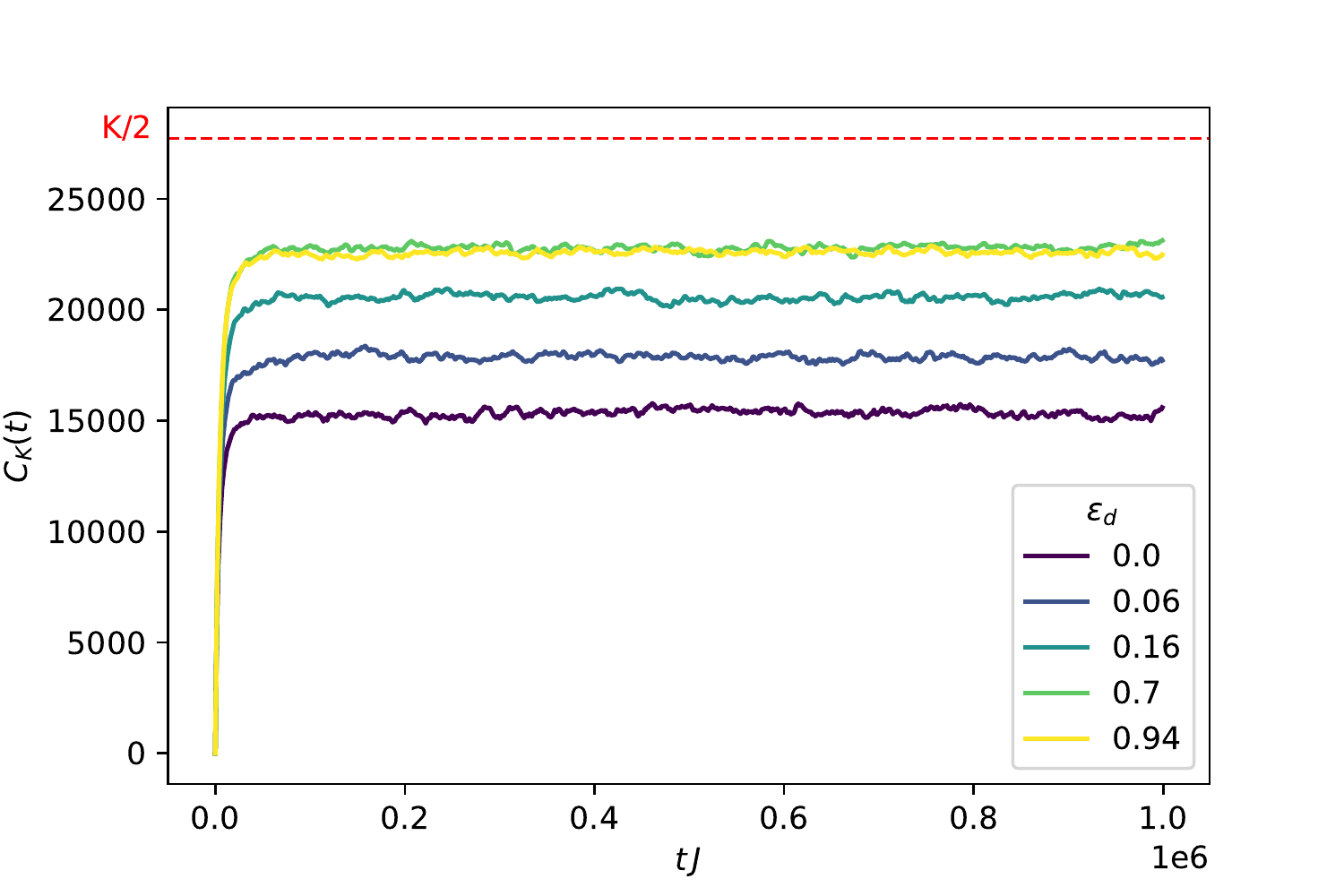}
        \caption{}
        \label{fig:}
    \end{subfigure}
    \caption{Results for the time-dependent profile of K-complexity at increasing time scales, for the same system of Fig. \ref{fig:N11M6Hd}.  The Hamiltonian is $H_{XXZ}+\epsilon_d H_d$ with $N=11$ spins in the $M=6, P=+1$ sector with $J_{zz}=1.1$, and the operator is $\mathcal{O}=S^z_{5}+S^z_{7}$ with trace removed. The Krylov space dimension for this setup is $K=55461$ which equals the upper bound for the Krylov space dimension. The final plot shows the saturation values of K-complexity, with the value of $K/2$ shown for reference.}
    \label{fig:KC_time}
\end{figure}

Indeed, we find consistently that the saturation value of K-complexity is affected by the strength of the integrability breaking term, and generally increases with the value of the integrability breaking coefficient, as can be seen in Figures \ref{fig:KC_sat_XXZ_Hd} and \ref{fig:KC_sat_XXZ_NNN}. The late-time saturation value of K-complexity as a fraction of the Krylov space dimension can be read off from the vertical lines in the figures, where the $x$-axis was scaled according to the corresponding Krylov space dimension. Another interesting aspect is the time-dependent profile of K-complexity at various time scales and for different integrability breaking strength, for which results are presented in Figure \ref{fig:KC_time}.  Again we find a consistent relationship between the strength of the integrability breaking term and the value of K-complexity.

\section{RMT results and dependence on universality class}\label{Sect:RMT}

This Section gathers results on the saturation value of K-complexity for different operators in Random Matrix Theory, to be used as a reference to compare with the results obtained in the chaotic regime of the deformed XXZ Hamiltonian studied in Sections \ref{sec:XXZ_rstats} and \ref{sect_XXZ_KCsat}.

We shall consider systems with a Hilbert space of dimension $D$, equipped with a random Hamiltonian $H$ drawn from a Gaussian ensemble, with probability measure
\begin{equation}
    \centering
    \label{RMT_Gaussian_measure}
    p(H) DH = \exp\left\{-\frac{D}{2 \sigma^2}\text{Tr}\left(H^\dagger H\right)\right\}DH\,,
\end{equation}
where $DH$ is a flat measure, the standard deviation $\sigma$ sets the energy units (and was set to $1$ in the numerics), and $H$ is a complex hermitian or a real symmetric matrix depending on whether we work with the Gaussian Unitary Ensemble (GUE) or with the Gaussian Orthogonal Ensemble (GOE), respectively\footnote{The third canonical Gaussian ensemble, which we don't study in this article, is the Gaussian Symplectic Ensemble (GSE). It addresses time-reversal-invariant fermionic systems displaying Kramer's degeneracy.} \cite{Haake_Book}.

\subsection{Influence of the structure of the seed operator}

A detailed numerical study reveals that the behavior of K-complexity, and in particular its late-time saturation value, is not only controlled by the statistics of the Hamiltonian spectrum, but also influenced by the structure of the operator under consideration. As an extreme illustration of this, Appendix \ref{appx_FlatOp} shows analytically that an operator that is constant in the energy basis, which is a very atypical observable in any system, features a late-time K-complexity saturation value of $\sim\frac{K}{2}$ regardless of the spectrum of the underlying Hamiltonian. In contrast, a typical operator in RMT should satisfy the \textit{RMT operator Ansatz} (see e.g. \cite{DAlessio:2015qtq}) for its matrix elements in the energy basis:
\begin{equation}
    \centering
    \label{RMT_op_Ansatz_maintext}
    \langle E_a | \mathcal{O} | E_b \rangle = O \delta_{ab} + \frac{1}{\sqrt{D}}r_{ab} ~,
\end{equation}
where all $\left\{r_{ab}\right\}$ are independent random numbers\footnote{In fact, only those $r_{ab}$ with $a\geq b$ are independent, as the rest are determined from the latter if the operator is hermitian.} drawn from a normal distribution with zero mean and unit variance; they are either real or complex depending on the universality class at hand. The one-point function term $O$ in (\ref{RMT_op_Ansatz_maintext}) will not be important for the current analysis because, as explained in Appendix \ref{appx_Connected}, we shall work with traceless operators.

\begin{figure}
    \centering
    \includegraphics[width=7.5cm]{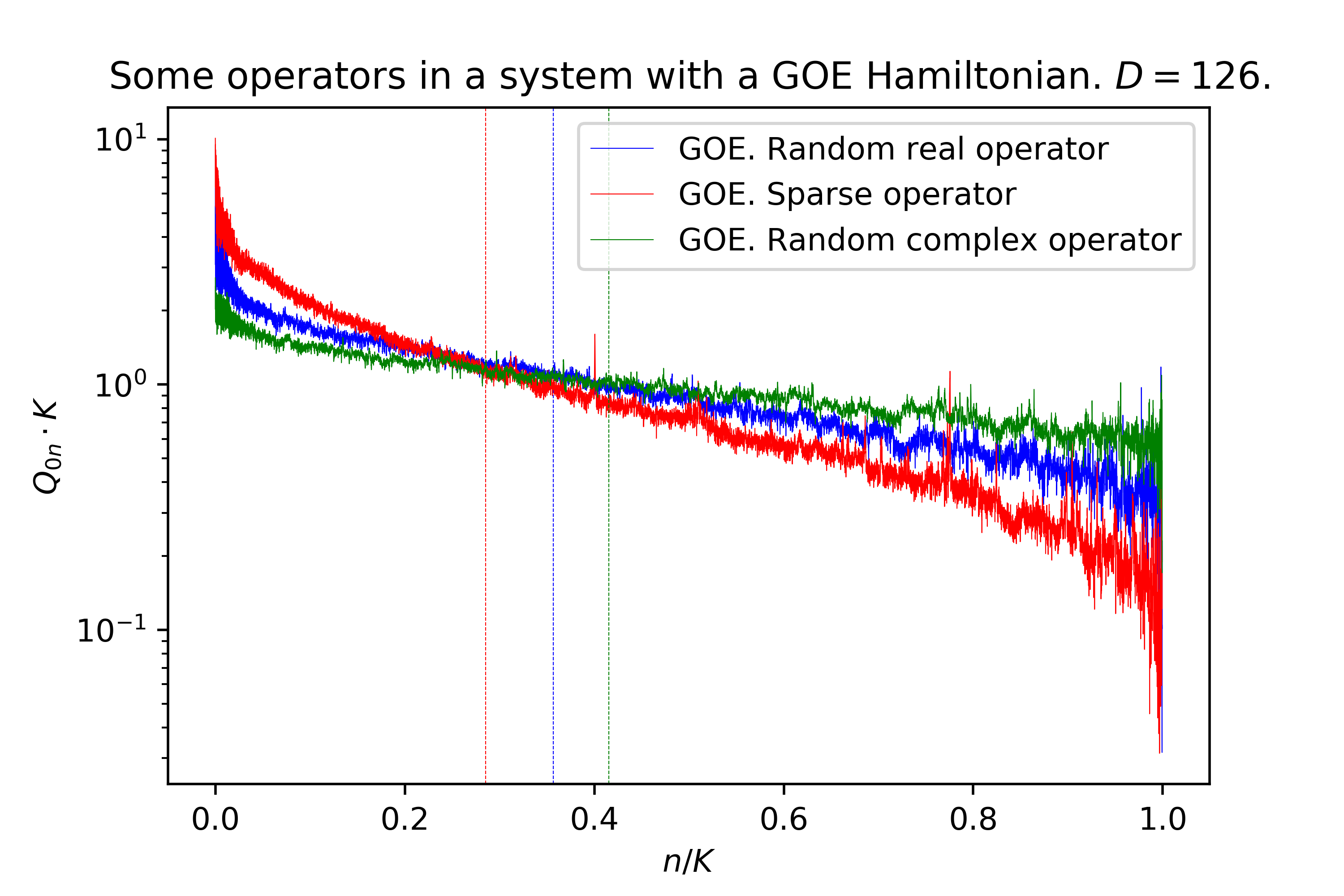} \includegraphics[width=7.5cm]{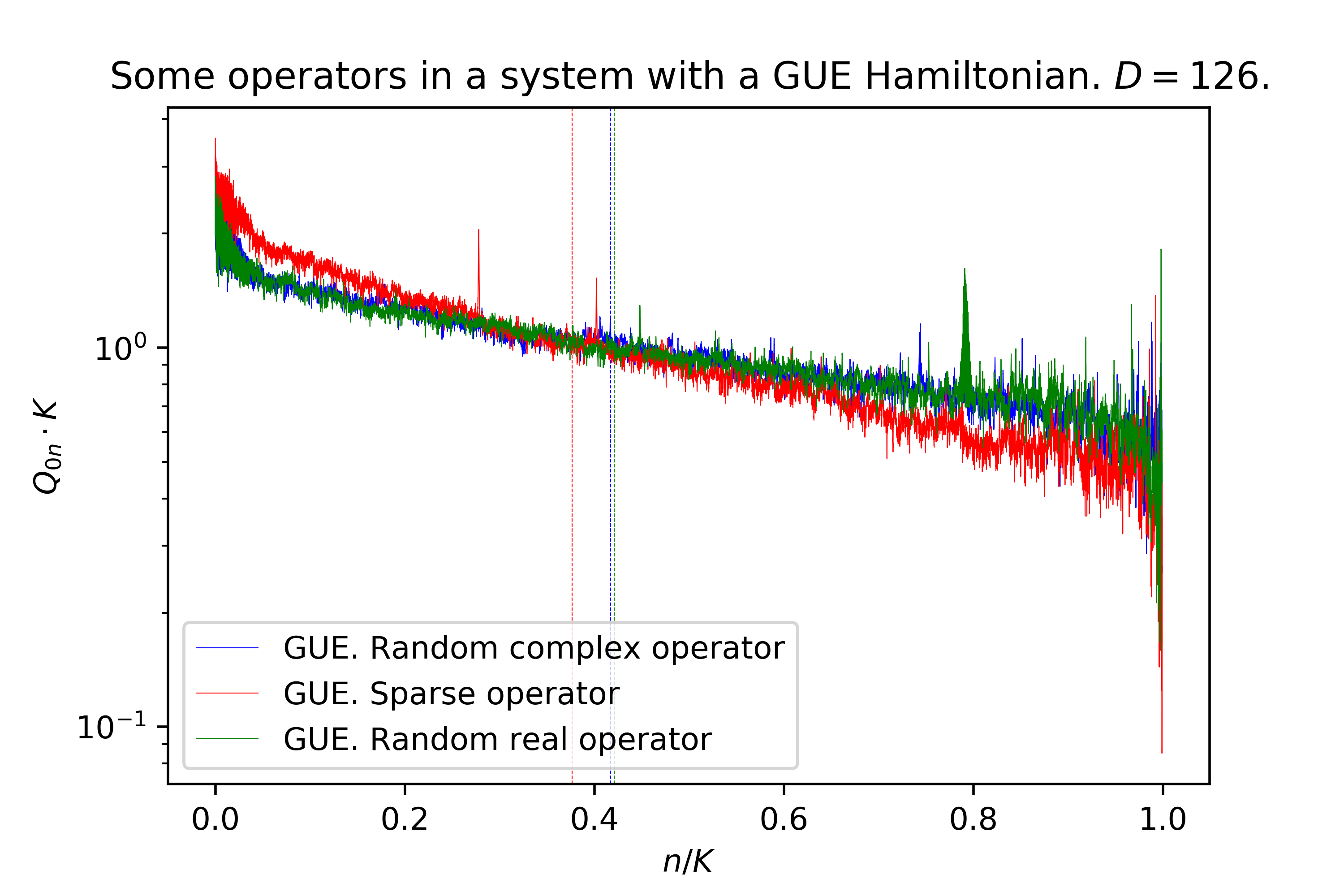}
    \caption{Long-time averaged operator wave packet and saturation value of K-complexity for different operator choices and Hamiltonians drawn from two RMT ensembles. K-complexity saturation values are marked by vertical lines. The Hilbert space dimension chosen was $D=126$, and the obtained Krylov dimension saturates the upper bound \cite{Rabinovici:2020ryf}, verifying $K=D^2-D+1=15751$. \textbf{Left:} GOE Hamiltonian. Operator choices: a sparse operator in the basis in which the Hamiltonian is drawn, a random real operator drawn from a Gaussian distribution in the same basis, and a random complex operator again drawn in the same basis. We observe that the random operators saturate at higher values as compared to the sparse operator, and within them, the one that breaks time-reversal (i.e. the complex one) has the highest complexity saturation value. \textbf{Right:} Same choices of operator, but for a Hamiltonian drawn from GUE. In this case time reversal is anyway broken by the Hamiltonian itself, which is why the two random operators have quantitatively very similar features, both having a complexity saturation value slightly higher than that of the sparse operator.}
    \label{fig:RMT_SomeOps}
\end{figure}

Operators satisfying the Ansatz (\ref{RMT_op_Ansatz_maintext}) can be constructed as sparse operators in the basis in which the Hamiltonian is drawn from the Gaussian ensemble, or as random matrices with independent entries. In both cases, the change-of-basis matrix that brings the operator to the energy basis is a random unitary drawn from the Haar measure and for sufficiently large $D$ they both agree with the structure (\ref{RMT_op_Ansatz_maintext}). Results on the late-time behavior of K-complexity for both operator choices in the different universality classes can be found in Figure \ref{fig:RMT_SomeOps}, which suggests that the saturation value of K-complexity is sensitive to the universality class to which the Hamiltonian belongs as well as to the choice of operator and, in particular, to whether the operator breaks time-reversal or not. Note that, in general, the complexity saturation values are below $\frac{K}{2}$.

\subsection{Deviations from the RMT Ansatz: ETH operators}\label{subsect:RMT_ETH}

\begin{figure}
    \centering
    \includegraphics[width=12cm]{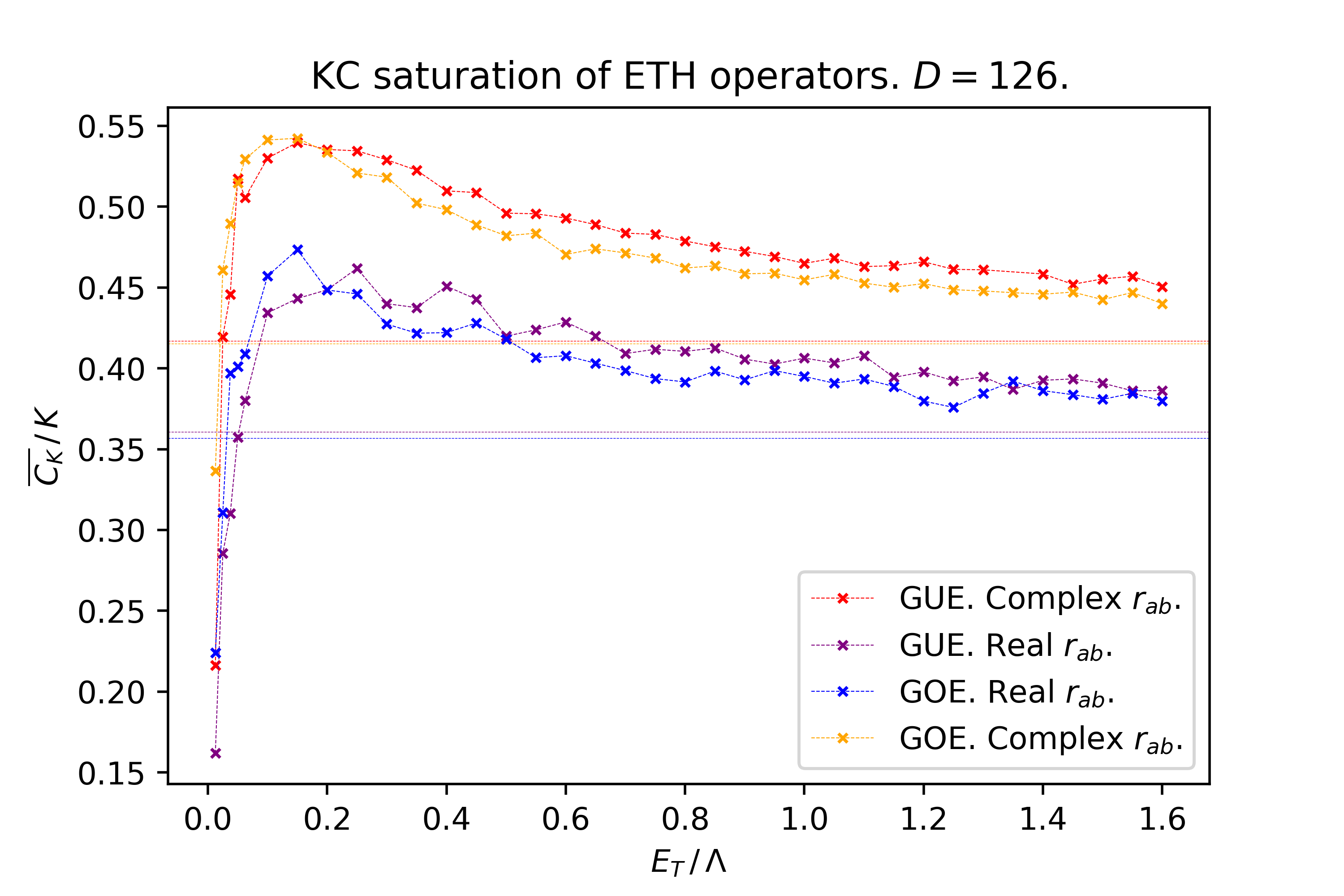}
    \caption{K-complexity saturation value as a function of the Thouless energy for different ETH operators (either real or complex in the energy basis) with Hamiltonians drawn from two different RMT ensembles (GOE and GUE). The horizontal lines mark the asymptotic value for $E_T\to\infty$, corresponding to the case of observables satisfying the pure RMT operator Ansatz. In order to mod out system-dependent scaling of the energy spectrum due to the choice of normalization of the Hamiltonian, the Thouless energy was normalized by the total bandwith $\Lambda$, allowing for potential comparisons with other systems.}
    \label{fig:Csat_Vs_Thouless}
\end{figure}

In \cite{Rabinovici:2020ryf} we studied complex SYK$_4$, which is a chaotic system with richer features than just RMT and displayed a complexity saturation value close to $\frac{K}{2}$.
Two features may be regarded as responsible for that behavior: the Wigner-Dyson statistics satisfied by the Hamiltonian spectrum, and the fact that the operators studied satisfy the eigenstate thermalization hypothesis (ETH) \cite{Deutsch_1991,Srednicki:1994mfb,DAlessio:2015qtq,Sonner:2017hxc,Nayak:2019khe}, which is an extension of the RMT Ansatz (\ref{RMT_op_Ansatz_maintext}) that accounts for a smoothly varying density of states:

\begin{equation}
    \centering
    \label{ETH_Ansatz}
    \langle E_a | \mathcal{O} | E_b \rangle = O(\overline{E})\delta_{ab}+e^{-\frac{S(\overline{E})}{2}}\,f_{\mathcal{O}}(\overline{E},\omega)\,r_{ab}~,
\end{equation}
where $r_{ab}$ are independent (up to hermiticity), identically distributed normal random variables with zero mean and unit variance, and $\overline{E}\equiv(E_a+E_b)/2$ and $\omega \equiv E_a-E_b$ are, respectively, the average energy and the energy difference between the corresponding levels. $O(\overline{E})$ and $S(\overline{E})$ are the microcanonical one-point function and entropy, respectively, and the function $f_{\mathcal{O}}(\overline{E},\omega)$ gives the Fourier transform of the connected two-point function, sometimes denoted \textit{spectral function} \cite{Parker:2018yvk}. Disregarding the $\overline{E}$-dependence, the high-frequency tails of this function are known to be bounded from above by an exponential profile:
\begin{equation}
\centering
\label{f_bound}
    f_{\mathcal{O}}(\overline{E},\omega)\lesssim e^{-\frac{\omega}{E_T}},
\end{equation}
where $E_T$ is the Thouless energy, which is itself constrained by a system-dependent upper bound, and controls the regime of applicability of RMT. For the sake of the current analysis, we generated operators following the Ansatz (\ref{ETH_Ansatz}) where the $\overline{E}$-dependence was taken to be constant and the $\omega$-dependence was chosen to saturate the bound (\ref{f_bound}) with an adjustable Thouless energy. The (rescaled) off-diagonal elements in the energy basis $\left\{ r_{ab} \right\}$ were chosen to be either real or complex. The saturation value of K-complexity as a function of the Thouless energy for the different choices of Hamiltonian and operator are depicted in Figure \ref{fig:Csat_Vs_Thouless}. The different choices of Hamiltonian and operator can be classified according to how they comport regarding time reversal. If we define the time reversal transformation $\mathcal{T}$ as an anti-unitary transformation that acts as complex conjugation $\mathcal{T}\overset{*}{=} K$ in the basis in which the Hamiltonian is drawn from the Gaussian ensemble, we can make the following identifications:

\begin{itemize}
    \item GOE + real $r_{ab}$: This situation matches that of a time-reversal preserving operator in a system with a Hamiltonian that preserves $\mathcal{T}$, as they both are real in the computational basis\footnote{For simplicity, here we refer to the basis in which the Hamiltonian is drawn from the corresponding ensemble as the \textit{computational basis}.}, and therefore the operator will still be real in the energy basis.
    \item GOE + complex $r_{ab}$: This case describes the situation in which the Hamiltonian is $\mathcal{T}$-invariant but the operator is not. The operator matrix elements in the computational basis will be complex and, since the change-of-basis matrix for going to the energy basis is a real orthogonal matrix, it will also have complex entries in the energy basis.
    \item GUE + complex $r_{ab}$: Since the Hamiltonian already breaks time reversal, the matrix of eigenvectors expressed in coordinates over the computational basis will be a random unitary, and hence in general the operator will have complex entries in the energy basis regardless of whether it was real or complex in the computational basis (i.e. regardless of whether it is invariant under $\mathcal{T}$ or not, respectively.)
    \item GUE + real $r_{ab}$: Along the lines of the previous point, we shall conclude that this configuration is just an atypical case, not particularly physically meaningful for discussions regarding time reversal. We have nevertheless still kept the results for this case in Figure \ref{fig:Csat_Vs_Thouless} for the sake of completeness of the analysis.
\end{itemize}

Disregarding for the discussion the seemingly unphysical GUE+real configuration, Figure \ref{fig:Csat_Vs_Thouless} illustrates the fact that in systems where time-reversal is broken, either by the Hamiltonian or by the operator, depict a systematically higher K-complexity saturation value. We have also observed a continuous dependence of the saturation value on the Thouless energy that can pump up the former from the lower limiting value attained when the observable satisfies the pure RMT operator Ansatz throughout the spectrum.

\subsection{ETH in the deformed XXZ}

As the integrability-breaking defects studied in Section \ref{sec:XXZ_rstats} are made stronger, the spectrum of the Hamiltonian of the deformed XXZ chain transitions from Poissonian statistics to Wigner-Dyson statistics. At the same time, it is possible to see that the seed operator under consideration transitions from a non-ETH regime when $\epsilon_d$ is small to having and ETH structure when $\epsilon_d$ attains the value that makes the spectrum of the Hamiltonian chaotic. This phenomenon was already studied in works like \cite{Rigol2020,LeBlond:2019eoe,Rigol_LeBlod2020}.

\begin{figure}
    \centering
    \includegraphics[width=7.5cm]{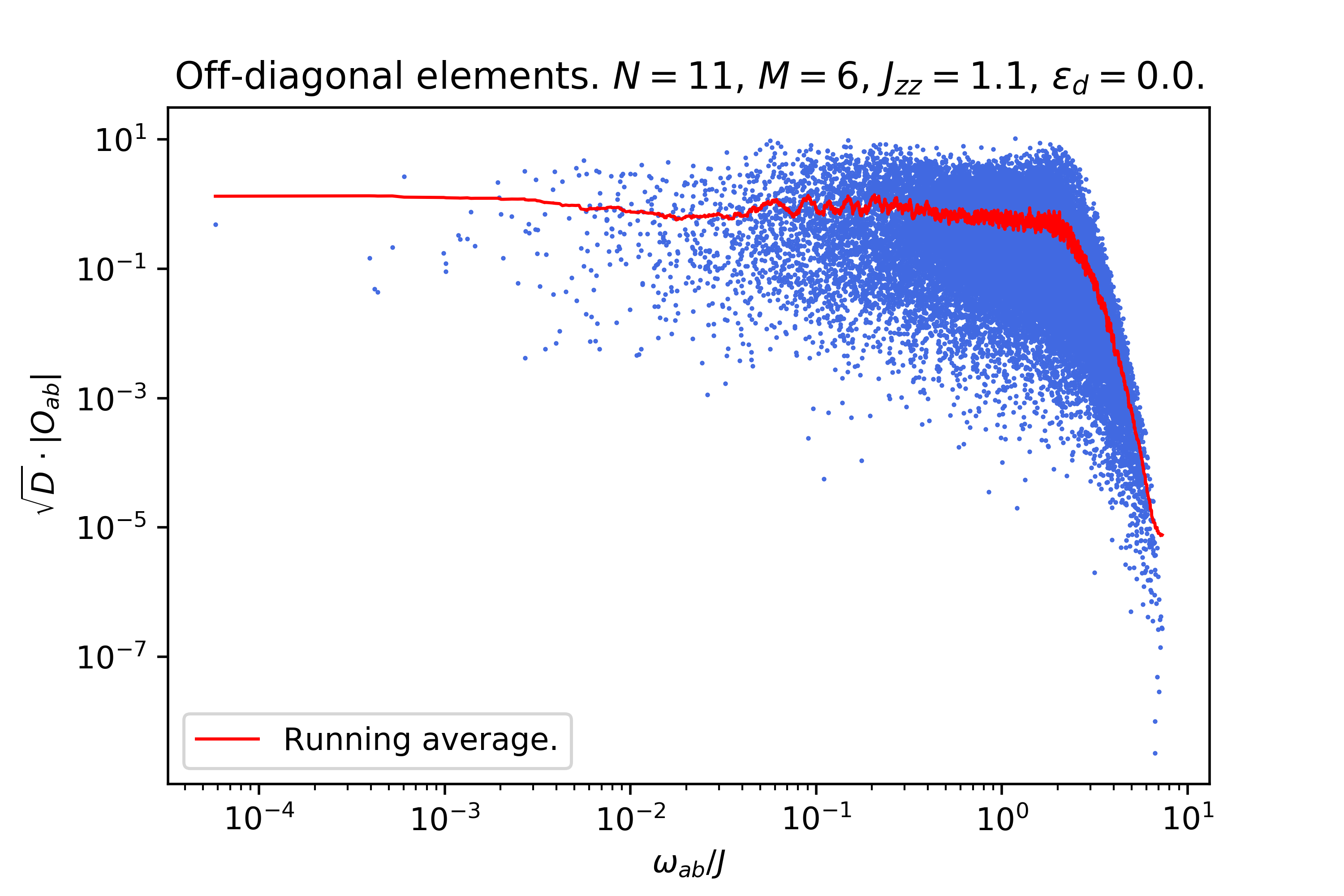} \includegraphics[width=7.5cm]{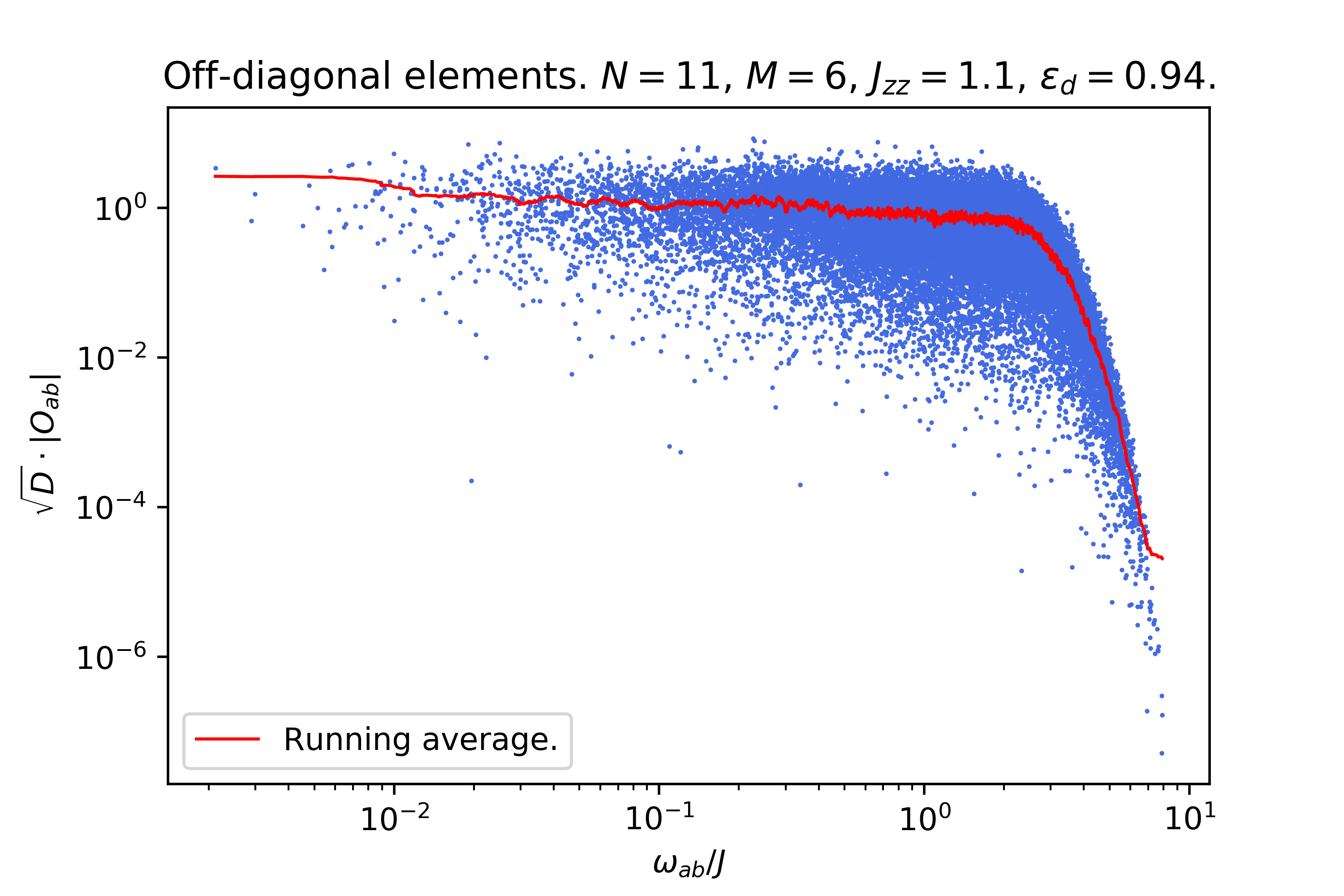} \\
    \includegraphics[width=7.5cm]{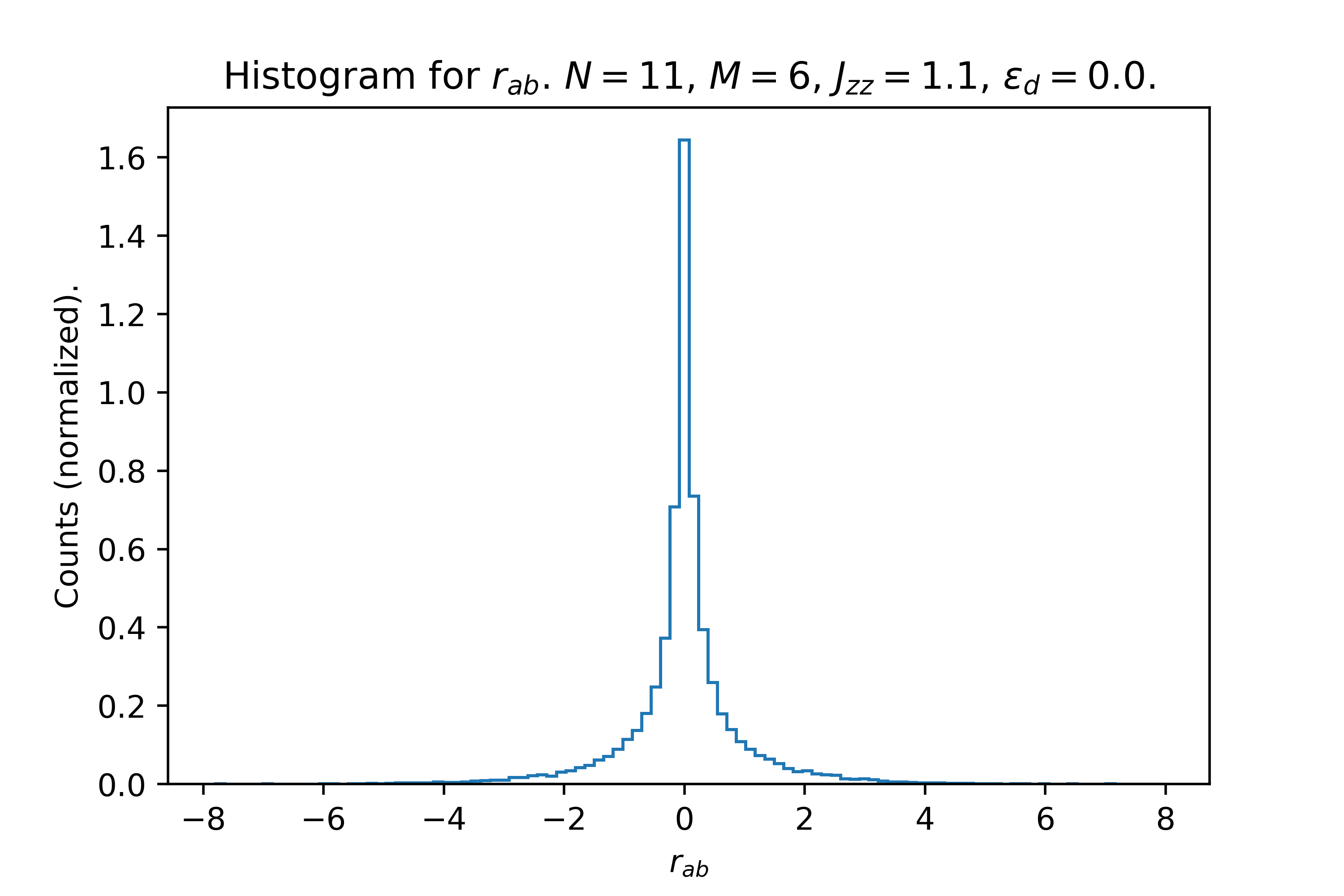} \includegraphics[width=7.5cm]{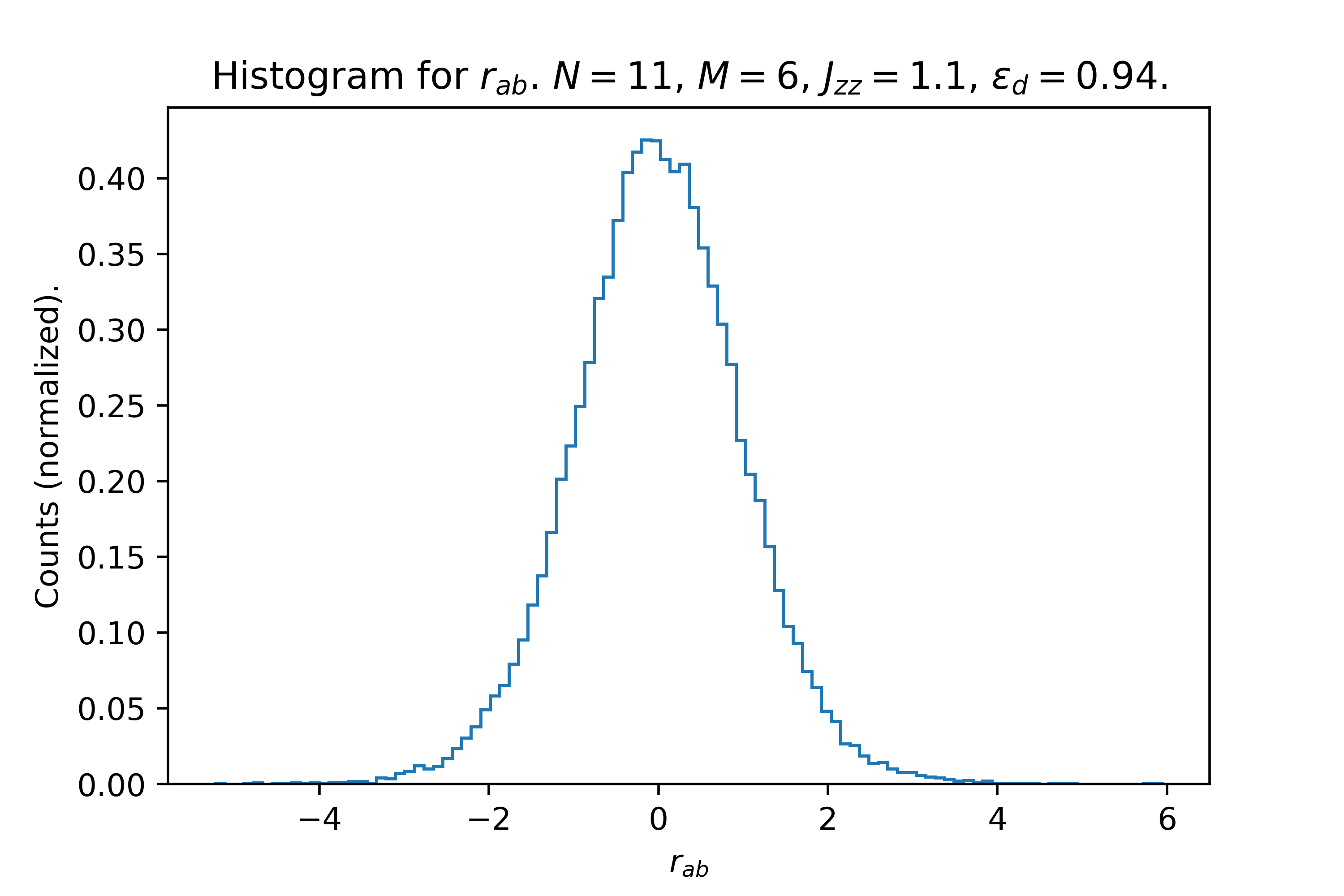} \\
    \includegraphics[width=7.5cm]{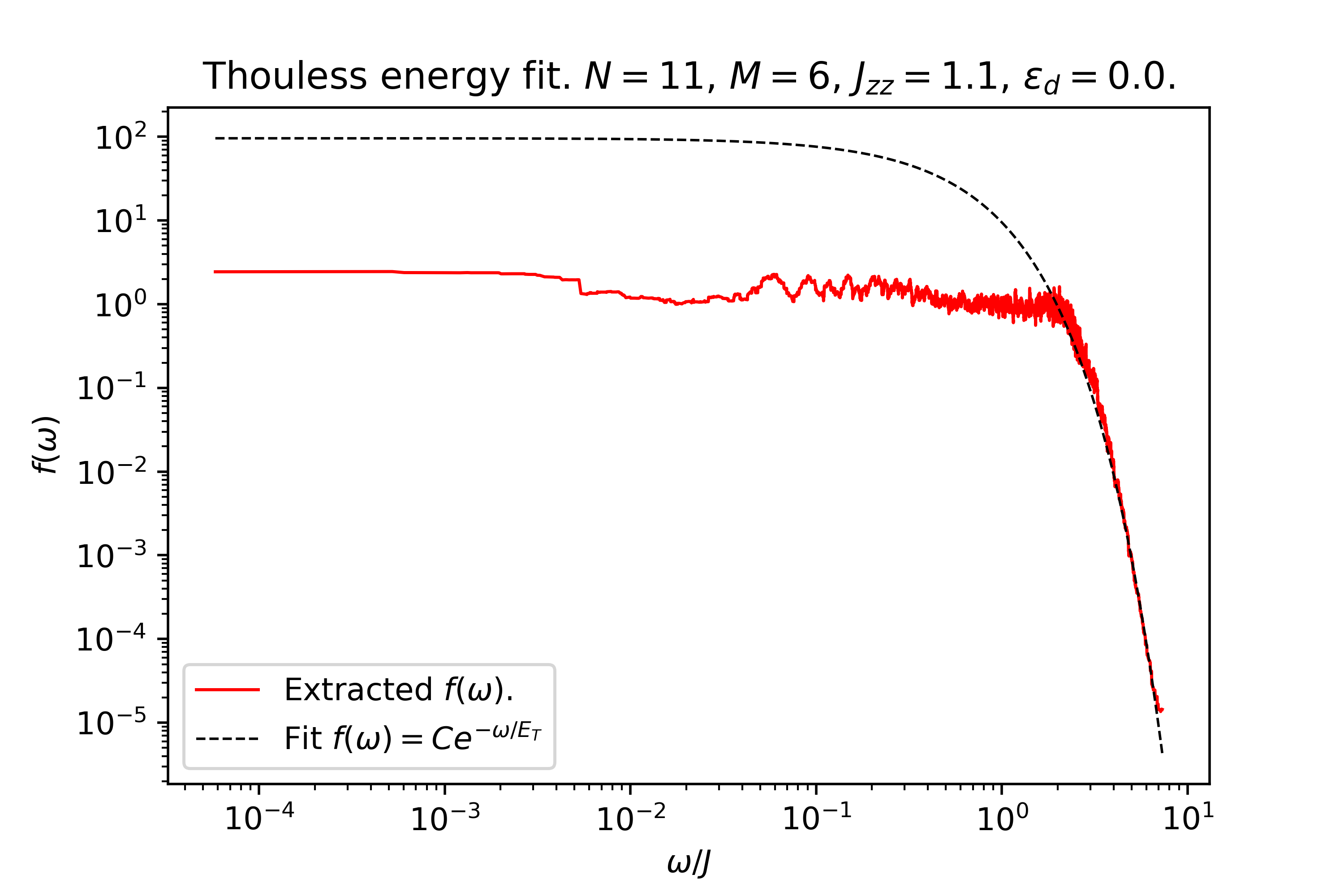} \includegraphics[width=7.5cm]{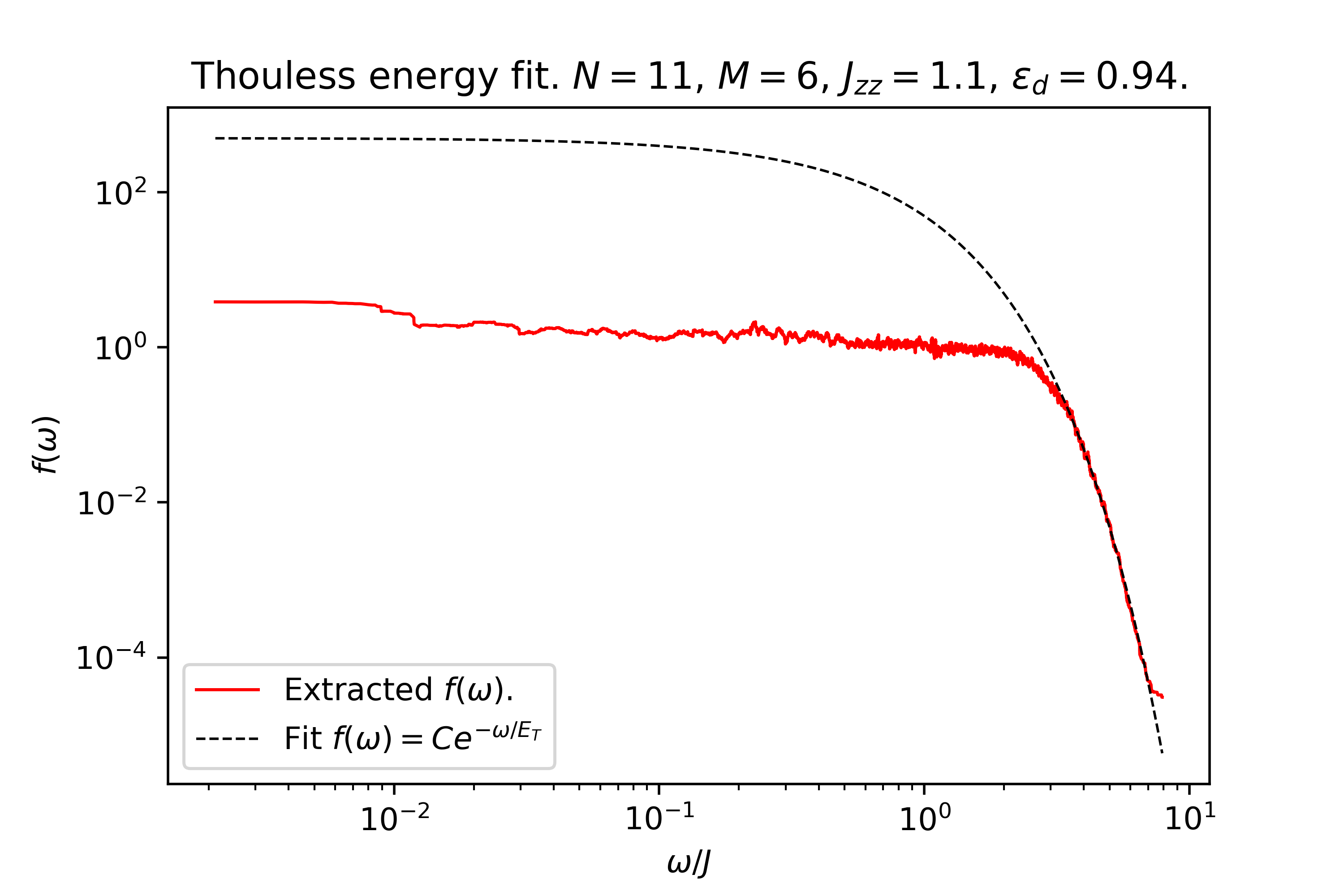}
    \caption{ETH checks for the system and operator studied in Figure \ref{fig:N11M6Hd}. On the \textbf{left} we show the integrable regime $\epsilon_d=0$ while on the \textbf{right} we consider the chaotic regime with $\epsilon_d=0.94$. Since we are studying operators with a zero one-point function, the ETH check can focus on just the off-diagonal elements of the operator in the energy basis. We note, in agreement with previous works (such as \cite{Rigol2020,LeBlond:2019eoe,Rigol_LeBlod2020}), that the difference between the integrable and the chaotic phase is subtle: in both cases it is possible to extract a smooth envelope $f(\omega_{ab})$ for the off-diagonal matrix elements of the operator as a function of the energy difference $\omega_{ab}\equiv E_a-E_b$, whose high-frequency tail can be fitted by an exponential form $f(\omega)=C e^{-\omega/E_T}$ yielding in both cases a very similar Thouless energy $E_T/\Lambda\sim 0.055$ (the r-value of the fit being around $0.992$), where $\Lambda$ denotes the spectral bandwidth. The difference between the integrable and the chaotic case is that in the former the fluctuations of the off-diagonal elements are not quite Gaussian, and hence they cannot be claimed to fulfill ETH, whereas they become more Gaussian in the latter, where the integrability-breaking defect is stronger. }
    \label{fig:ETH_XXZ}
\end{figure}

Here we present results on ETH checks for two extreme values of $\epsilon_d=0,\,0.94$ for the system and operator that were analyzed in Figure \ref{fig:N11M6Hd}. Figure \ref{fig:ETH_XXZ} displays the result. In the integrable regime, the operator does not fulfill the ETH Ansatz because the fluctuations are not Gaussian. In the chaotic regime ($\epsilon_d=0.94$) the operator is seen to agree with the ETH Ansatz displaying a Thouless energy normalized by the spectral bandwidth of roughly $\frac{E_T}{\Lambda}\sim 0.05$. In this chaotic regime, we have that the spectrum of the Hamiltonian is chaotic and that the operator fulfills the ETH Ansatz with a certain Thouless energy; since these are precisely the only two ingredients defining the systems studied in Subsection \ref{subsect:RMT_ETH} and depicted in Figure \ref{fig:Csat_Vs_Thouless}, one can compare the K-complexity saturation values. The universality class at hand for our deformed XXZ is ``GOE+real'', and we note from Figure \ref{fig:Csat_Vs_Thouless} that indeed, for a Thouless energy satisfying $\frac{E_T}{\Lambda}\sim 0.05$ one expects a K-complexity saturation value around $0.4K$, consistent with what we found in Figure \ref{fig:N11M6Hd} when $\epsilon_d=0.94$.

\section{Discussion}\label{Sect:Disc}
We have explored the behavior of K-complexity of a strongly coupled integrable model with an integrability breaking deformation both at the integrable point and in the chaotic phase. The purpose of this study is to delineate what kind of behavior this notion of complexity has in a chaotic system as opposed to a strongly coupled integrable one. We have found further strong evidence for the picture proposed in \cite{Rabinovici:2020ryf,Rabinovici:2021qqt} namely that an exponentially large K-complexity saturation value at late times is a generic feature of a quantum chaotic system, while integrable systems, even strongly coupled ones which do not show exact degeneracies of energy levels, have quantitatively lower saturation values. We studied K-complexity and its late-time saturation value for XXZ systems with two types of integrability breaking terms and found that increasing the value of the coefficient of the integrability breaking term, results in an increasing value for the late-time saturation of K-complexity. We further compared the late-time saturation value in the chaotic regime to the results for RMT in the corresponding universality class, finding reasonable agreement. Along the way, we noted that non-zero operator one-point functions can influence the late-time behavior of K-complexity in a similar fashion to how the disconnected piece of the two-point function governs the late-time regime of the correlator if it is not subtracted.  

We will end this discussion with a number of open questions and further avenues of research surrounding K-complexity. Firstly, it would clearly be of great interest to develop a more analytical understanding of the non-trivial phenomena we have uncovered in this paper, perhaps by attacking it from the angle of `Krylov localization' as in \cite{Rabinovici:2021qqt}, that is to analytically establish localization of the relevant part of the wave-function on the Krylov chain. A further interesting avenue to explore concerns time-dependent Hamiltonians, and how one might develop a viable generalization of Krylov complexity in such circumstances\footnote{Work in progress with J.L.F. Barb\'{o}n.}. Finally, since a particularly interesting application of K-complexity is in the context of holographic duality, it is desirable to incorporate K-complexity into the holographic dictionary. In this context it is intriguing to remark that \cite{Basteiro:2022zur} recently proposed a discrete holographic dual of the aperiodic XXZ chain. It may be of interest to generalize our results to this case, to allow a more direct comparison to a simple discrete holographic code.

\acknowledgments{The numerical computations were performed on the Landau cluster at the Hebrew University and on the Baobab HPC cluster at the University of Geneva. This work has been supported in part by the Fonds National Suisse de la Recherche Scientifique (Schweizerischer Nationalfonds zur Förderung der wissenschaftlichen Forschung) through Project Grants 200020 182513 and the NCCR 51NF40-141869 The Mathematics of Physics (SwissMAP).  The work of ER and RS is partially supported by the Israeli Science Foundation Center of Excellence.}

\appendix

\section{Connected part of autocorrelation function and saturation value of K-complexity}\label{appx_Connected}
This Appendix analyses the impact of the operator one-point function on the saturation value of K-complexity at late times. 
Let us first remind how the operator one-point function dominates the two-point function plateau. We shall do so by assuming that the operator satisfies the Eigenstate Thermalization Hypothesis (ETH).
Consider a hermitian normalized operator whose elements in the energy basis are given by
\begin{eqnarray}
\label{Op_spectral_decomp}
    \mathcal{O}&=&\sum_{a,b=1}^D  O_{ab}|E_a\rangle \langle E_b|
\end{eqnarray}
where $D$ is the Hilbert space dimension. Note that in (\ref{Op_spectral_decomp}) the operator matrix elements in the energy basis are defined such that $O_{ab}=\langle E_a | \mathcal{O} | E_b \rangle$.
With this convention, the ETH Anstatz takes the usual form; suppressing energy dependence in the matrix elements of the Ansatz (since we are interested in order-of-magnitude estimates), it boils down to the RMT operator Ansatz:
\begin{equation}
    \centering
    \label{RMT_op_Ansatz}
    O_{ab} = O\delta_{ab}+\frac{1}{\sqrt{D}}r_{ab},
\end{equation}
where $O$ gives (up to non-perturbative corrections) the one-point function of the operator and is taken not to scale with $D$, and the matrix $(r_{ab})$ is drawn from a Gaussian ensemble with unit variance (and hence the elements $r_{ab}$ are also of order $D^0$). Note that the operator (\ref{Op_spectral_decomp}) with the matrix elements given by (\ref{RMT_op_Ansatz}) is normalized\footnote{By this, we mean that the norm of the operator whose matrix elements satisfy (\ref{RMT_op_Ansatz}) doesn't scale with $D$.} according to the operator inner product
\begin{eqnarray}
\label{Inner_prod_norm_one}
    \|\mathcal{O}\|^2 = \frac{1}{D} \mathrm{Tr}\Big[\mathcal{O}^\dagger \mathcal{O} \Big] = \frac{1}{D}\sum_{a,b=1}^D |O_{ab}|^2 = 1.
\end{eqnarray}
The autocorrelation function is given by
\begin{eqnarray}
\label{Two-pt-function}
    \phi_0(t)&=&\big\langle \mathcal{O}^\dagger \mathcal{O}(t)\big\rangle=\frac{1}{D} \mathrm{Tr}\Big[\mathcal{O}^\dagger \mathcal{O}(t) \Big] = \frac{1}{D} \sum_{a,b=1}^D |O_{ab}|^2 e^{i(E_a-E_b)t} ~.
\end{eqnarray}
Due to normalization of the operator (\ref{Inner_prod_norm_one}), the two-point function (\ref{Two-pt-function}) starts at $1$, i.e. $\phi_0(0)=1$. The Ansatz (\ref{RMT_op_Ansatz}) has some implications on the late-time behavior of $\phi_0(t)$, which we can study by performing a long-time average:

\begin{equation}
    \centering
    \label{Two-Pt-LongTAvg}
    \overline{\phi_0}:=\lim_{T\to+\infty}\frac{1}{T}\int_0^T dt\, \phi_0(t).
\end{equation}
We can now use that, for $\omega\neq 0$:
\begin{equation}
    \centering
    \label{Phases_average_out}
    \lim_{T\to +\infty}\frac{1}{T}\int_{0}^T dt\, e^{i\omega t}=\lim_{T\to +\infty}\frac{1}{T}\left. \left[ \frac{e^{i\omega t}}{i\omega} \right] \right|_{t=0}^T=\frac{1}{i\omega}\lim_{T\to+\infty}\frac{e^{i\omega T}-1}{T}=0.
\end{equation}
With this, assuming no exact degeneracies in the energy spectrum, the long-time average of (\ref{Two-pt-function}) eliminates the contribution of the off-diagonal matrix elements and yields:
\begin{equation}
    \centering
    \label{Two-Pt_LongT_Diagonal}
    \overline{\phi_0}=\frac{1}{D}\sum_{a=1}^D|O_{aa}|^2=\frac{1}{D}\sum_{a=1}^D \left( O + \frac{r_{aa}}{\sqrt{D}} \right)^2= O^2 + \frac{2O}{D^{3/2}}\sum_{a=1}^D r_{aa} + \frac{1}{D^2}\sum_{a=1}^D r_{aa}^2,
\end{equation}
where in the second equality we have used the Ansatz (\ref{RMT_op_Ansatz}). We thus conclude that the long-time average of the two-point function is dominated by the square of the one-point function. This fact is also in qualitative agreement with large-N factorization, i.e. in the thermodynamic limit two-point function becomes disconnected at late times. 

In order to probe spectral correlations we can choose to subtract explicitly the one-point function squared from the auto-correlation function (\ref{Two-pt-function}), which defines the so-called connected two-point function:

\begin{equation}
    \centering
    \label{Two-pt-connected}
    \phi_0^{(c)}(t):= \Bigg\langle \bigg( \mathcal{O} - \langle \mathcal{O} \rangle\bigg)\bigg(\mathcal{O}(t)-\langle\mathcal{O}\rangle\bigg) \Bigg\rangle = \big\langle \mathcal{O}\mathcal{O}(t) \big\rangle-\big\langle \mathcal{O} \big\rangle^2,
\end{equation}
where, to alleviate notational crowding, we have implicitly assumed that $\mathcal{O}$ is hermitian, and we have made use of the fact that the one-point function is time-independent, $\big\langle \mathcal{O}(t) \big\rangle = \big\langle \mathcal{O} \big\rangle$. Again, making use of the expression of the operator $\mathcal{O}$ in the energy basis, we can write (\ref{Two-pt-connected}) as:
\begin{equation}
    \centering
    \label{Two_Pt_Conn_Ebasis}
    \phi_0^{(c)}(t) = \frac{1}{D}\sum_{a,b=1}^D|O_{ab}|^2 e^{it(E_a-E_b)}\,-\,\frac{1}{D^2}\sum_{a,b=1}^DO_{aa}O_{bb}, 
\end{equation}
where we have used that $\langle \mathcal O \rangle = \frac{1}{D}\text{Tr}[\mathcal O]$ is the infinite-temperature one-point function of the operator $\mathcal{O}$. As defined in (\ref{Two-pt-connected}), the connected two-point function is not normalized so that its value at $t=0$ is exactly one, but this is not important because we can still prove that $\phi_0^{(c)}(t=0)$ is of order one, i.e. its value doesn't scale with $D$:

\begin{equation}
    \centering
    \label{Two_Pt_Conn_t0}
    \phi_0^{(c)}(t=0)=\frac{1}{D}\sum_{a,b=1}^D |O_{ab}|^2 - \frac{1}{D^2}\sum_{a,b=2}^D O_{aa}O_{bb} = \frac{1}{D^2}\sum_{a,b=1}^D |r_{ab}|^2-\frac{1}{D}\left\{ \frac{1}{D^2}\sum_{a,b=1}^Dr_{aa}r_{bb} \right\}, 
\end{equation}
as can be seen plugging in the Ansatz (\ref{RMT_op_Ansatz}). We note that the leading term in (\ref{Two_Pt_Conn_t0}) is the first term of the last expression, consisting of a sum of $D^2$ numbers of order one, divided by a $D^2$ factor, and hence $\phi_0^{(c)}(t=0)$ is a number of order $D^0$.

Now, we can estimate the height of the late-time plateau by computing the long-time average of the connected two-point function:
\begin{equation}
    \centering
    \label{Two_Pt_Conn_LongT_Avg}
    \overline{\phi_0^{(c)}}= \lim_{T\to \infty} \frac{1}{T}\int_0^T dt \, \phi_0^{(c)}(t) = \frac{1}{D}\sum_{a=1}^D O_{aa}^2 - \frac{1}{D^2}\sum_{a,b=1}^D O_{aa} O_{bb}.
\end{equation}
Plugging the Ansatz (\ref{RMT_op_Ansatz}) in (\ref{Two_Pt_Conn_LongT_Avg}) we again find that the terms involving the order-one quantity $O$ cancel out, yielding:
\begin{equation}
    \centering
    \label{Two_Pt_Conn_LongT_Avg_result}
    \overline{\phi_0^{(c)}}= \frac{1}{D}\left\{ \frac{1}{D}\sum_{a=1}^D r_{aa}^2 + \frac{1}{D^2}\sum_{a,b=1}^D r_{aa}r_{bb} \right\}.
\end{equation}
The quantity inside the braces is of order $D^{0}$. We thus conclude that the connected two-point function has a long-time average of order $\frac{1}{D}$, and that this is deduced from the ETH-like Ansatz (\ref{RMT_op_Ansatz}). To argue that $\phi_0^{(c)}(t)$ actually plateaus at $\frac{1}{D}$, one should prove that its long time variance is (exponentially) suppressed\footnote{We shall refer to quantities of order $\frac{1}{D}$ or smaller as \textit{exponentially suppressed} because the Hilbert space dimension is typically exponential in the number of degrees of freedom $S$ of the system, i.e. $D\sim e^{S}$.}, so that the function remains close to its long-time average at late times. We shall do that later, when studying the long time average of the square of the two-point function. But before that, we can note that there was a simpler way to derive the previous results, by defining a new operator $\widetilde{\mathcal{O}}$ obtained by subtracting the one-point function from the initial operator $\mathcal{O}$:
\begin{equation}
    \centering
    \label{Connected_Op}
    \widetilde{\mathcal{O}}=\mathcal{O}-\mathbb{1}\langle \mathcal{O} \rangle = \mathcal{O}-\frac{1}{D}\text{Tr}[\mathcal{O}]\mathbb{1}.
\end{equation}
Using the operator Ansatz for $\mathcal{O}$ given in (\ref{RMT_op_Ansatz}), we note that the matrix elements of $\widetilde{\mathcal{O}}$ in the energy basis are given by:
\begin{equation}
    \centering
    \label{Matrix_elements_otilde}
    \widetilde{O}_{ab} = \frac{1}{\sqrt{D}}\widetilde{r}_{ab},
\end{equation}
where all $\widetilde{r}_{ab}$ are of order one and the matrix $\widetilde{R}\equiv \left( \widetilde{r}_{ab} \right)$ is related to the matrix $R\equiv \left( r_{ab} \right)$ through:
\begin{equation}
    \centering
    \label{Rtilde_vs_R}
    \widetilde{R} = R - \mathbb{1} \langle R \rangle.
\end{equation}
In particular:
\begin{equation}
    \centering
    \label{rtilde_vs_r}
    \widetilde{r}_{ab} = r_{ab}-\frac{\delta_{ab}}{D}\sum_{c=1}^D r_{cc},
\end{equation}
from where it is apparent that all $\widetilde{r}_{ab}$ are of order one and that they follow the exact constraint $\text{Tr}[\widetilde{R}]=\sum_{a=1}^D\widetilde{r}_{aa}=0$.

This operator redefinition is useful because we immediately note that the connected two-point function of $\mathcal{O}$ is identically equal to the full two-point function of $\widetilde{\mathcal{O}}$:

\begin{equation}
    \centering
    \label{Two_Pt_Conn_Tilde}
    \phi_0^{(c)}(t) = \big\langle \widetilde{\mathcal{O}} \widetilde{\mathcal{O}}(t) \big\rangle.
\end{equation}
And thus, using the expression of $\widetilde{O}$ in the energy basis (\ref{Matrix_elements_otilde}) it is straightforward to see that:
\begin{equation}
    \centering
    \label{Two_Pt_Otilde}
    \phi_0^{(c)}(t) = \frac{1}{D}\sum_{a,b=1}^D |\widetilde{O}_{ab}|^2 e^{it(E_a-E_b)} = \frac{1}{D^2}\sum_{a,b=1}^D |\widetilde{r}_{ab}|^2 e^{it(E_a-E_b)},
\end{equation}
from where it is immediate that $\phi_0^{(c)}(t=0)\sim D^0$ and that $\overline{\phi_0^{(c)}}\sim \frac{1}{D}$.

A similar argument can now be applied to transition probabilities on the Krylov chain.
The long-time average of the transition probability is given by (\ref{Q0n_LTA})
\begin{eqnarray}
    Q_{0n} := \overline{|\phi_n|^2} = \lim_{T\to \infty} \frac{1}{T} \int_0^T |\phi_n(t)|^2 dt,
\end{eqnarray}
which for $n=0$ takes the form (\ref{Q00_LTA}).
From the Ansatz (\ref{RMT_op_Ansatz}) we find that:
\begin{eqnarray}
     \frac{1}{D^2}\sum_{a,b=1}^D |O_{aa}|^2 |O_{bb}|^2 \sim O^4 + \mathit{O}\left( \frac{1}{D} \right).
\end{eqnarray}
\begin{eqnarray}
    \frac{1}{D^2}\sum_{a\neq b=1}^D  |O_{ab}|^4 \sim \mathit{O}\left(\frac{1}{D^2}\right).
\end{eqnarray}
We thus find that the long-time-average of the square of the autocorrelation function behaves like
\begin{eqnarray} \label{large_phi0}
     \overline{|\phi_0|^2} &\sim& O(1)+ O\left(\frac{1}{D}\right).
\end{eqnarray}
The time averaged transition probability $Q_{00}$ takes an order-one value controlled by the one-point function. In order to see an exponentially suppressed plateau, we again need to work with the connected two-point function $\phi_0^{(c)}(t)$, and the associated probability $P_{00}^{(c)}(t):=\phi_0^{(c)}(t)^2$ (note that the two-point function is always real provided that the operator is hermitian, which we assume), whose long-time average we shall denote $Q_{00}^{(c)}$. As we showed in (\ref{Two_Pt_Conn_t0}), $\phi_0^{(c)}(t=0)\sim 1$, and therefore $P_{00}^{(c)}(t=0)\sim 1$. Likewise, $Q_{00}^{(c)}$ can be expressed in terms of the matrix elements of the traceless operator $\widetilde{\mathcal{O}}$:
\begin{equation}
    \centering
    \label{Q00_Conn_otilde}
    Q_{00}^{(c)}=\frac{1}{D^2}\sum_{a,b=1}^D\left[ \widetilde{O}_{aa}^2\widetilde{O}_{bb}^2 + |\widetilde{O}_{ab}|^4 \right] = \frac{1}{D^4}\sum_{a,b=1}^D \left[ \widetilde{r}_{aa}^2\widetilde{r}_{bb}^2 + |\widetilde{r}_{ab}|^4 \right] \sim \frac{D^2}{D^4}=\frac{1}{D^2}.
\end{equation}
And hence $P_{00}^{(c)}(t)$ plateaus\footnote{Actually, in order to show that $P_{00}^{(c)}(t)$ plateaus at the long-time average $Q_{00}^{(c)}$, we should also show that the long-time variance of $P_{00}^{(c)}(t)$ is suppressed, otherwise a strongly oscillating function could still be compatible with the long-time average prediction. This proof is doable (even though cumbersome), but $P_{00}^{(c)}(t)\sim e^{-2S}$ at late times seems to hold for ETH operators in chaotic systems like cSYK$_4$ according to numerical checks.} at late times at $\frac{1}{D^2}\sim e^{-2S}$. Incidentally, note that $Q_{00}^{(c)}$ gives the long-time variance of $\phi_0^{(c)}(t)$, and hence showing that (\ref{Q00_Conn_otilde}) is suppressed concludes the proof that the connected two-point function is close to the plateau value at late times, as anticipated above.

\subsection{Example: Hopping vs number operator in cSYK$_4$} \label{app:SYK}

In a previous work on cSYK$_4$ \cite{Rabinovici:2020ryf}, we studied the K-complexity of hopping operators, $h_{ij}=c_i^\dagger c_j + h.c$. These operators have a zero one-point function, and hence their two-point function is connected. However, non-universal effects due to a non-zero one-point function can be probed if we consider, for example, an on-site number operator $n_i=c_i^\dagger c_i$. Indeed, since in cSYK$_4$ we work in fixed occupation sectors, the one-point functions of the on-site number operators are constrained by the relation:

\begin{equation}
    \centering
    \label{Number_ev}
    N=\sum_{i=1}^L n_i\;\Longrightarrow\;\langle N \rangle = \sum_{i=1}^L \langle n_i \rangle, 
\end{equation}
where $N$ is the total number operator and $L$ is the number of sites (or rather, the number of complex fermions). Hence, (\ref{Number_ev}) together with the fact that $\langle n_i \rangle \geq 0$ implies that at least one on-site number operator needs to have a non-zero expectation value whenever $\langle N \rangle > 0$. In fact, the chaotic character of cSYK$_4$ seems to distribute equally the expectation value accross all the $n_i$, and given a fixed occupation sector we can thus estimate $\langle n_i \rangle \sim \frac{\langle N \rangle}{L}$ for all $i=1,...,L$, i.e. the one-point function equals the filling ratio.

\begin{figure}
    \centering
    \includegraphics[width=12cm]{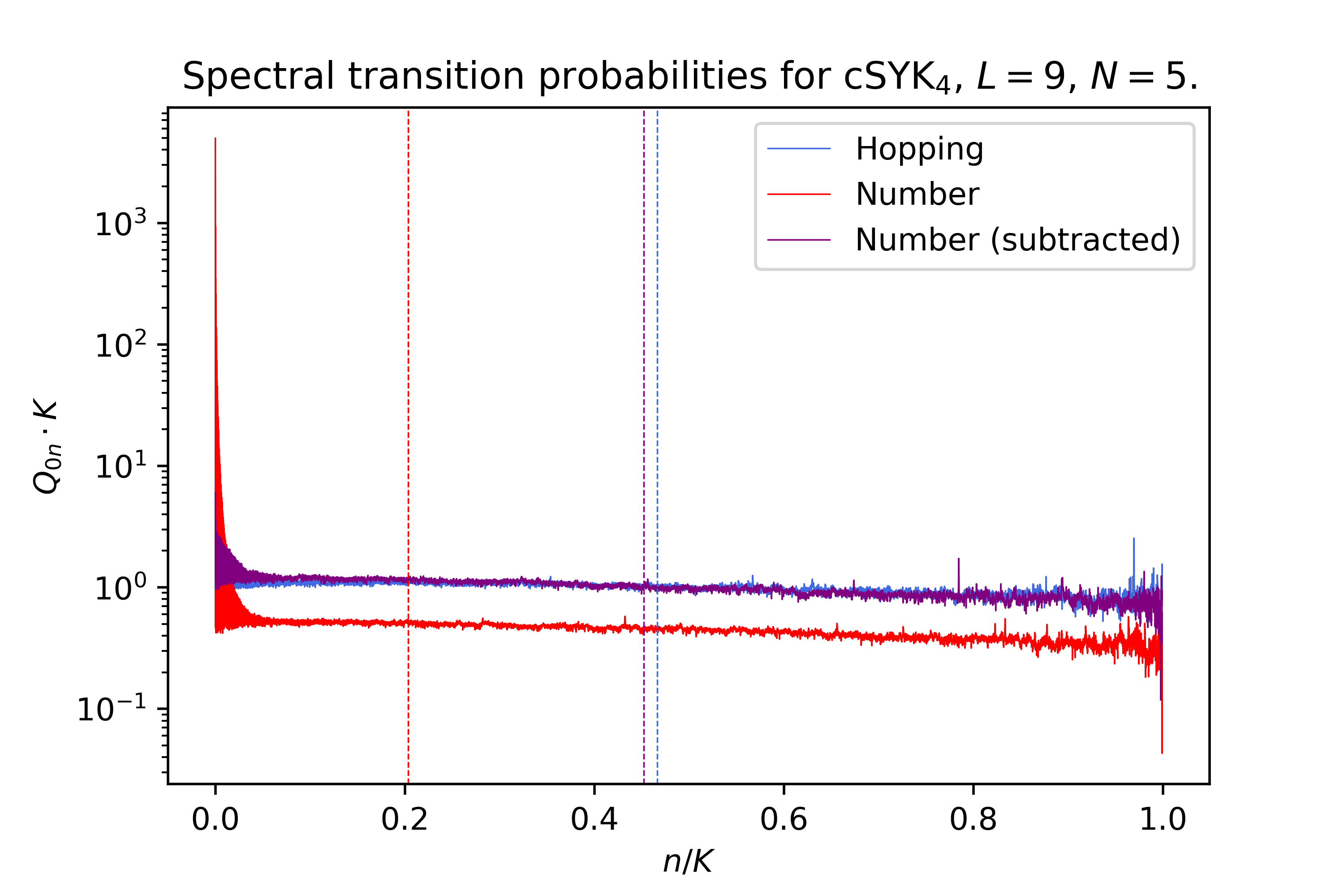}
    \caption{Averaged transition probabilities $Q_{0n}$ in cSYK$_4$ for the hopping operator and both the full and the subtracted version of the number operator. Vertical lines mark the estimated K-complexity saturation value. For this system size, $K=15751\sim 10^4$, and we observe that $Q_{00}\sim 1$ for the full number operator, signaling that indeed the one-point function dominates the late-time regime of the transition probability. The rest of the $Q_{0n}$ with $n>1$ for the full number operator seem to be rather uniformly distributed, but with a lower value due to the constraint $\sum_{n=0}^{K-1}Q_{0n}=1$. This eventually enforces a complexity saturation value of $\sim 0.2K$, much below the naively expected $\sim\frac{K}{2}$, which the hopping operator does display. Conversely, both for the hopping operator and for the subtracted number operator, which have a zero one-point function by construction, all the transition probabilities are rather uniformly distributed around $\frac{1}{K}$, yielding a K-complexity saturation value much closer to $\frac{K}{2}$.}
    \label{fig:Transition_Prob_SYK_Ops}
\end{figure}

This non-zero one-point function controls the averaged transition probability $Q_{00}$, which becomes of order one in system size and, as discussed above, has the effect of lowering the K-complexity saturation value in (\ref{KC_LTA}). Figure \ref{fig:Transition_Prob_SYK_Ops} illustrates this claim.

In order to avoid this, we can seed the Lanczos algorithm with the subtracted version of the operator, $\widetilde{\mathcal{O}}$, which by construction has a zero one-point function. Following the usual arguments \cite{Recursion_Method, Parker:2018yvk}, we find that the Lanczos coefficients $\widetilde{b}_n$ of $\widetilde{\mathcal{O}}$ are in one-to-one correspondence with the moments $\widetilde{\mu}_n$ of the connected two-point function of $\mathcal{O}$, $\phi_0^{(c)}(t)$, and that the K-complexity long-time average is given by:

\begin{equation}
    \centering
    \label{KC-longT-otilde}
    \overline{C_K}=\sum_{n=0}^{K-1} n \widetilde{Q_{0n}},
\end{equation}
where $\widetilde{Q_{00}}=Q_{00}^{(c)}$. Therefore, as argued above, for an ETH operator this last quantity will be exponentially suppressed, hence not competing with the other uniformly distributed $\widetilde{Q_{0n}}$ with $n>0$ and allowing for a K-complexity saturation value closer to $K/2$. This is illustrated in Figure \ref{fig:Transition_Prob_SYK_Ops}.

\subsection{Role of the one-point function in XXZ} \label{app:XXZ}
\begin{figure}
    \centering
    \includegraphics[width=12cm]{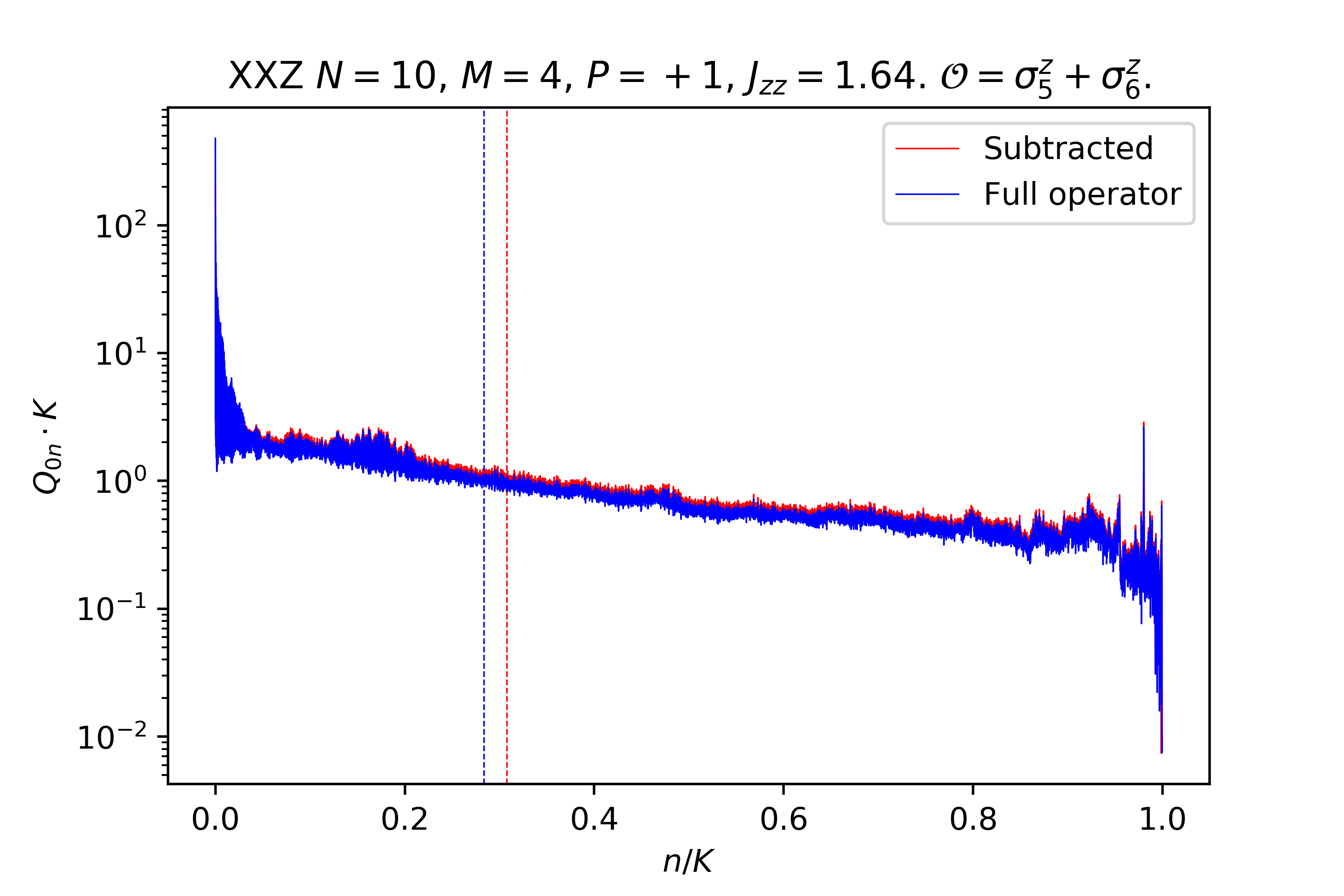}
    \caption{Long-time averaged transition probabilities for an instance of XXZ studied in \cite{Rabinovici:2021qqt}, this time also considering a version of the operator where the non-vanishing one-point function has been subtracted. In contrast to the cSYK$_4$ case, this time the subtraction of the one-point function doesn't alter drastically the K-complexity saturation value, since in this case undersaturation is due to the monotonously decaying profile of $Q_{0n}$ that we associated to Anderson localization on the Krylov chain in \cite{Rabinovici:2021qqt}, together with the fact that even the connected part of the two-point function is itself not exponentially suppressed at late times due to the integrable nature of the system.}
    \label{fig:XXZ_Connected_VS_Full}
\end{figure}

This may raise some concern regarding previous work in XXZ \cite{Rabinovici:2021qqt}, as the operators used in that case were on-site Pauli sigma matrices, whose one-point function in fixed-magnetization Hilbert space sectors are also constrained by the value of the total magnetization in the given sector, through:

\begin{equation}
    \centering
    \label{Total_Mag_ev}
    S^z = \frac{1}{2}\sum_{n=1}^N \sigma_n^z\;\longrightarrow\;\langle S^z \rangle = \frac{1}{2}\sum_{n=1}^N\langle \sigma_n^z \rangle ,
\end{equation}
where, for XXZ, $N$ denotes the number of chain sites. Since XXZ is integrable, we don't necessarily assume that all $\langle \sigma_n^z \rangle$ are similar, but it is anyway clear from (\ref{Total_Mag_ev}) that in general they need not be zero. This might make one think that the XXZ calculations should be re-made taking as an input the subtracted version of on-site Pauli matrices. Figure \ref{fig:XXZ_Connected_VS_Full} illustrates that, in this case, subtracting the one-point function doesn't alter qualitatively the results because the connected part of the two-point function in this integrable system is already not exponentially suppressed, and hence the late-time value of the two-point function doesn't change drastically if one subtracts the disconnected part from it.

\section{Profile of transition probability for flat operator}\label{appx_FlatOp}
Consider a dense operator with constant matrix elements $O_{ab}=1$ for all $a,b=1,\dots,D$. In the Krylov basis such an operator has the following profile
\begin{eqnarray} \label{const_operator}
    |\mathcal{O}_0)=\Big(\underbrace{\frac{1}{D},\dots, \frac{1}{D}}_{\text{$\frac{D(D-1)}{2}$ terms} },\sqrt{\frac{1}{D}}, \underbrace{\frac{1}{D},\dots, \frac{1}{D}}_{\text{$\frac{D(D-1)}{2}$ terms} } \Big)
\end{eqnarray}
such that $\sum_{i=0}^{K-1}|O_i|^2=1$.  We will call such an operator a `flat' operator.

We recall from \cite{Rabinovici:2021qqt} that for an odd number of elements in the Krylov basis, which is  the case when no degeneracies are present and the operator has a non-zero projection over all Liouvillian frequencies (in such a case $K=D^2-D+1$ which is an odd number), the  Liouvillian eigenvector at the middle of its spectrum has zero eigenvalue 
\begin{eqnarray}
    \mathcal{L}|\omega_{\text{middle}}) = \omega_{\text{middle}}|\omega_{\text{middle}})=0 ~.
\end{eqnarray}
In general, the Liovillian eigenvectors can be expanded in the Krylov basis:
\begin{eqnarray}\label{omega_eigvec}
    |\omega_i) = \sum_{n=0}^{K-1} \psi_{ni}|\mathcal{O}_n)~.
\end{eqnarray}
For $|\omega_{\text{middle}})$ the coefficients $\psi_{n}$ satisfy
\begin{eqnarray}
    \psi_{2n} &=& (\mathcal{O}_{2n}|\omega_{\text{middle}}) = \psi_0 \prod_{i=1}^n \frac{b_{2i-1}}{b_{2i}} \equiv \psi_0 X_n \label{psi_even}\\
    \psi_{2n+1} &=& (\mathcal{O}_{2n+1}|\omega_{\text{middle}})  = 0~, \label{psi_odd}
\end{eqnarray}
where in (\ref{psi_even}) we defined $\prod_{i=1}^n \frac{b_{2i-1}}{b_{2i}} \equiv X_n$ for later convenience. 
Note that $\psi_0$ is determined by the middle element of (\ref{const_operator}), i.e. 
\begin{eqnarray} \label{psi0}
    \psi_0=\sqrt{\frac{1}{D}}
\end{eqnarray}
since in $\mathcal{L}$'s eigenvector matrix, the middle element of $|\mathcal{O}_0)$ is the first element in $|\omega_{\text{middle}}) $ as can be seen from (\ref{omega_eigvec}). 

\begin{figure}[H]
    \centering
    \begin{subfigure}[t]{0.45\textwidth}
    \centering
        \includegraphics[scale=0.45]{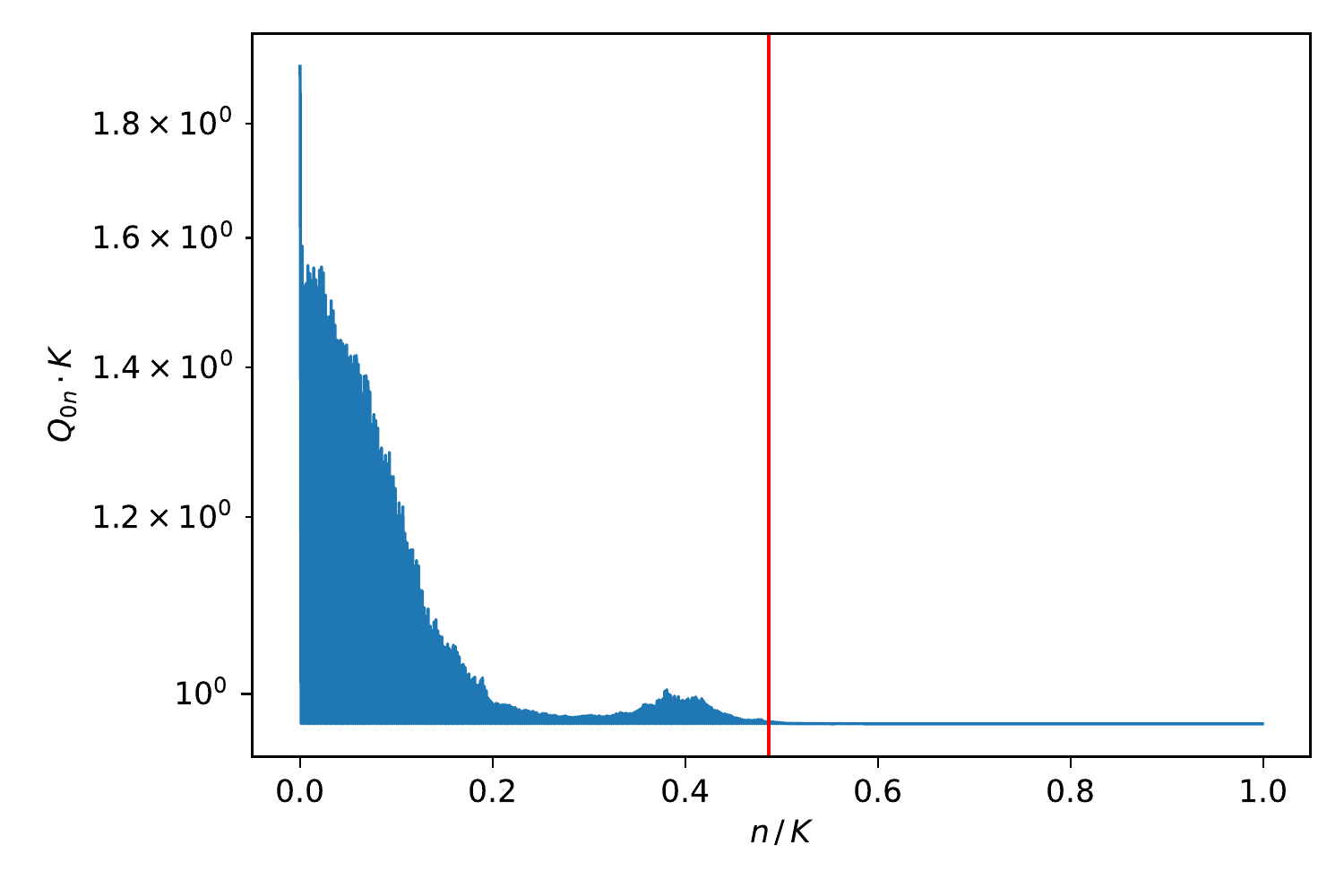}
        \caption{}
        \label{fig:}
    \end{subfigure}
    \hfill
    \begin{subfigure}[t]{0.45\textwidth}
    \centering
        \includegraphics[scale=0.45]{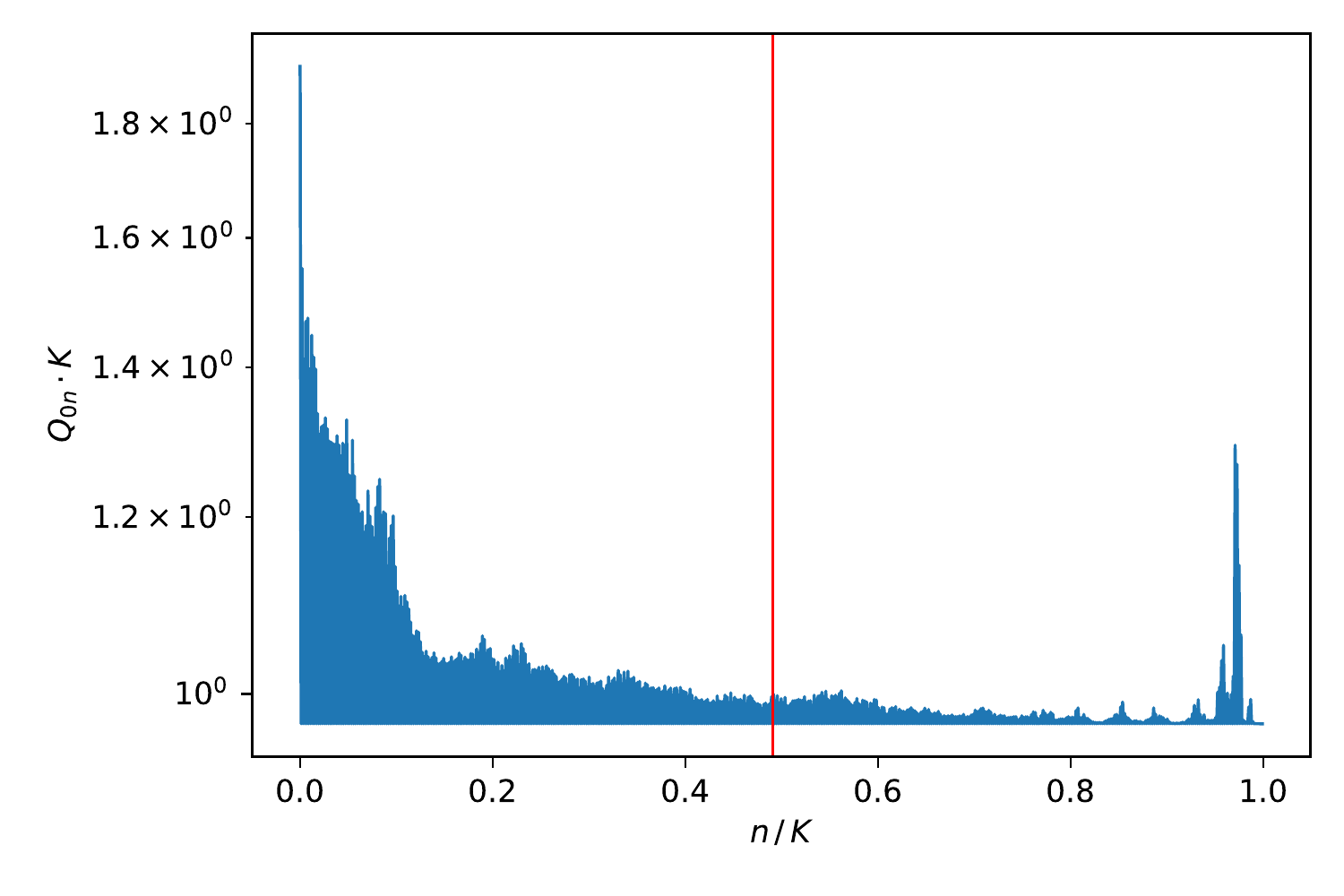}
        \caption{}
        \label{fig:}
    \end{subfigure}
    \caption{K-complexity saturation value for constant operator evolving under Hamiltonian taken from a GOE ensemble (left) and Hamiltonian with Poissonian statistics (right), both computed at $D=32$.  Note that both cases exhibit saturation value close to $K/2$.}
    \label{fig:GOE_vs_Poisson}
\end{figure}

With the information from (\ref{const_operator}), (\ref{psi_even}, \ref{psi_odd}) and  (\ref{psi0}) we can compute $Q_{0n}$ in terms of the Lanczos coefficients.
Starting with the definition (\ref{Q0n_LTA})
\begin{eqnarray}
    Q_{0n}=\sum_{i=0}^{K-1} |(\mathcal{O}_0|\omega_i)|^2 |(\mathcal{O}_n|\omega_i)|^2 
\end{eqnarray}
the first element is given by
\begin{eqnarray}
    Q_{00}=\sum_{i=0}^{K-1} |(\mathcal{O}_0|\omega_i)|^4 = D(D-1) \frac{1}{D^4} + \frac{1}{D^2} = \frac{2D-1}{D^3}~.
\end{eqnarray}
where we used (\ref{const_operator}) directly. The next element is
\begin{eqnarray}
    Q_{01}=\sum_{i=0}^{K-1} |(\mathcal{O}_0|\omega_i)|^2 |(\mathcal{O}_1|\omega_i)|^2 = \frac{1}{D^2} \sum_{i\neq \text{middle}} |(\mathcal{O}_1|\omega_i)|^2 + \frac{1}{D}\cdot 0 = \frac{1}{D^2}
\end{eqnarray}
where in the second equality we used the fact that the Krylov elements are normalized, hence $1=\sum_{i\neq \text{middle}} |(\mathcal{O}_1|\omega_i)|^2+|(\mathcal{O}_1|\omega_{\text{middle}})|^2$ and since from (\ref{psi_odd}), $(\mathcal{O}_1|\omega_{\text{middle}})=0$ we deduce that $\sum_{i\neq \text{middle}} |(\mathcal{O}_1|\omega_i)|^2=1$.

\begin{eqnarray}
    Q_{02} &=&\sum_{i=0}^{K-1} |(\mathcal{O}_0|\omega_i)|^2 |(\mathcal{O}_2|\omega_i)|^2 = \frac{1}{D^2} \sum_{i\neq \text{middle}} |(\mathcal{O}_2|\omega_i)|^2 + \frac{1}{D}  |(\mathcal{O}_2|\omega_{\text{middle}})|^2 \nonumber\\
    &=& \frac{1}{D^2}\left(1-\frac{1}{D}X_1^2\right) +\frac{1}{D}\left(\frac{1}{D}X_1^2\right) = \frac{1}{D^2} + \frac{X_1^2}{D^2}\left( 1-\frac{1}{D} \right) ~.
\end{eqnarray}
The rest of $Q_{0n}$ can be computed in a similar manner, and we conclude that for $n\geq 1$
\begin{eqnarray}
    Q_{0,2n} &=& \frac{1}{D^2} + \frac{X_n^2}{D^2}\left( 1-\frac{1}{D} \right) \\
    Q_{0,2n+1} &=& \frac{1}{D^2} ~.
\end{eqnarray}
One can check that this result is normalized
\begin{eqnarray}
    \sum_{n=0}^{K-1} Q_{0n} = \frac{2D-1}{D^3} +\left(\frac{K-1}{2}\right)\frac{1}{D^2}+\left(\frac{K-1}{2}\right)\frac{1}{D^2} +  \frac{D-1}{D^2} \sum_{n=1}^{\frac{K-1}{2}}\frac{X_n^2}{D} = 1
\end{eqnarray}
where we used $K=D^2-D+1$ and from normalization of $|\omega_{\text{middle}})$ we know that $\frac{1}{D}+\sum_{n=1}^{\frac{K-1}{2}} \frac{X_n^2}{D}=1$.
The value of K-complexity can then be estimated as follows:
\begin{eqnarray}
    \overline{C_K} &=& \sum_{n=0}^{K-1} n Q_{0n} = \frac{1}{D^2}\sum_{n=1}^{K-1} n + \frac{D-1}{D^2}  \sum_{n=1}^{\frac{K-1}{2}} 2n \frac{X_n^2}{D} \nonumber \\
    &=&\frac{1}{2}K(K-1)\frac{1}{D^2} + \frac{D-1}{D^2} {C_K}_{\text{middle}}
\end{eqnarray}
where ${C_K}_{\text{middle}}$ is the K-complexity of the eigenvector $|\omega_{\text{middle}})$\footnote{K-complexity for individual eigenvectors of the Liouvillian was defined in \cite{Rabinovici:2021qqt}.} and by definition $0\leq {C_K}_{\text{middle}}\leq K$. Hence it is found that for a flat operator
\begin{eqnarray}
    \frac{D^2}{2}-D+1-\frac{1}{2D} \leq \overline{C_K} \leq \frac{D^2}{2}-1+\frac{3}{2D}-\frac{1}{D^2}
\end{eqnarray}
which for large enough $D$ indicates that 
\begin{eqnarray}
    \overline{C_K} \sim \frac{D^2}{2} \sim \frac{K}{2}
\end{eqnarray}
independently of the spectrum or Lanczos coefficients data.  We show numerically that this is indeed the case by studying flat operators with Hamiltonians of GOE statistics and Poissonian statistics in Figure \ref{fig:GOE_vs_Poisson}.

\bibliography{references}

\end{document}